\documentclass[10pt,twocolumn,twoside]{IEEEtran}

\usepackage{cite}\usepackage{hyperref}
\usepackage{amsmath,amssymb,amsfonts}
\usepackage{algorithmic}
\usepackage{graphicx}
\usepackage{textcomp}
\usepackage{amsmath,amssymb,bm,bbm,mathrsfs,amscd}
\usepackage{calc}
\usepackage{color}
\usepackage{amsthm}
\usepackage{dsfont}
\usepackage{graphicx}
\usepackage{epstopdf}
\usepackage{epsfig}
\usepackage{tikz}
\usepackage{amsmath}
\usepackage{amsfonts}
\usepackage{amssymb}
\usepackage{rotating}
\usepackage{mathtools}
\usepackage{color}
\usepackage{subfig}
\usepackage{enumerate,pgfplots}
\newtheorem{theorem}{Theorem}
\newtheorem{definition}{Definition}
\newtheorem{proposition}{Proposition}
\newtheorem{lemma}{Lemma}
\newtheorem{corollary}{Corollary}
\newtheorem{example}{Example}
\newtheorem{remark}{Remark}
\newtheorem{assumption}{Assumption}
\newcommand{\ba}{\begin{array}}
\newcommand{\ea}{\end{array}}

\newcommand{\mb}{\boldsymbol}
\newcommand{\be}{\begin{equation}}
\newcommand{\ee}{\end{equation}}

\newcommand{\ds}{\displaystyle}

\newcommand{\eps}{\varepsilon}

\newcommand{\mc}{\mathcal}

\def\1{\boldsymbol{1}}
\def\etav{\boldsymbol{\eta}}

\newcommand{\R}{\mathbb{R}}

\newcommand{\nash}{\mc Y^*}
\newcommand{\rnash}{\mc Y^{\bullet}}
\newcommand{\xnash}{\mc X^*}
\newcommand{\xrnash}{\mc X^{\bullet}}
\newcommand{\xnashbal}{\mc X^*_{\etav}}
\newcommand{\xrnashbal}{\mc X^\bullet_{\etav}} 

\newcommand{\se}{\text{ if }}

\DeclareMathOperator*{\argmax}{argmax}
\DeclareMathOperator*{\argmin}{argmin}
\DeclareMathOperator{\sgn}{sgn}

\DeclareMathOperator{\dist}{dist}

\def\R{\mathbb{R}}

\def\f{{\mb f}}
\def\x{{\mb x}}
\def\y{{\mb y}}

\def\l{6.5}
\def\ll{2.7} 
\def\hh{1.3} 

\def\h{2.5} 

\def\BibTeX{{\rm B\kern-.05em{\sc i\kern-.025em b}\kern-.08em
    T\kern-.1667em\lower.7ex\hbox{E}\kern-.125emX}}
\title{Imitation dynamics in population games {on community networks}}
\author{Giacomo~Como,~\IEEEmembership{Member,~IEEE,}
        Fabio~Fagnani, and Lorenzo Zino
\thanks{Some of the results in the paper appeared in preliminary form in~\cite{cdc2017}.}
\thanks{G. Como and F. Fagnani are with the  Department of Mathematical Sciences ``G.L.~Lagrange,'' Politecnico di Torino, 10129 Torino, Italy  (e-mail: {\{giacomo.como;\,fabio.fagnani\}@polito.it}). L. Zino is with the  Faculty of Science and Engineering, University of Groningen, 9747 AG Groningen, The Netherlands  (e-mail: lorenzo.zino@rug.nl). G. Como is also with the Department of Automatic Control, Lund University, 22100 Lund, Sweden.}
\thanks{This work was partially supported by MIUR grant Dipartimenti di Eccellenza 2018--2022 [CUP: E11G18000350001], the Swedish Research Council [2015-04066], the Compagnia di San Paolo, the European Research Council [ERC-CoG-771687], and the Netherlands Organization for Scientific Research [NWO-vidi-14134].}%
}

\begin{document}

\maketitle

\begin{abstract}
We study the asymptotic behavior of deterministic, continuous-time imitation dynamics for population games over networks.  The basic assumption of this learning mechanism ---encompassing the replicator dynamics--- is that players {belonging to a single population} exchange information through pairwise interactions, whereby they get aware of the actions played by the other players and the corresponding rewards. {Using this information, they can revise their current action, imitating the one of the players they interact with.} The pattern of interactions regulating the learning process is determined by a community structure. First, the set of equilibrium points of such network imitation dynamics is characterized. Second, for the class of potential games {and for undirected and connected community networks, global asymptotic convergence is proved. In particular, our results guarantee convergence to a Nash equilibrium from every fully supported initial population state in the special case when the Nash equilibria are isolated and fully supported.} Examples and numerical simulations are offered to validate the theoretical results and counterexamples are discussed for scenarios when the assumptions on the community structure are not verified.
\end{abstract}


\IEEEpeerreviewmaketitle

\section{Introduction}

\IEEEPARstart{I}{n} the last decades, evolutionary game theory has emerged as a valuable mathematical paradigm to study the evolution of behaviors in social, economic, and biological network systems~\cite{Smith1982, Weibull1995, Hofbauer2003, Sandholm2010, Quijano2017}. Evolutionary game theory models these processes as the effect of a learning mechanism regulating the dynamics through which players in a population game revise their actions over time to improve their rewards.

In this paper, we focus on a class of learning mechanisms known as imitation dynamics~\cite{Weibull1995, Hofbauer2003}. Differently from other learning mechanisms, such as best-response dynamics or logit choice~\cite{Marden2012,Pavel2012}, imitation dynamics only require the players to have limited, {local} information on the structure of the game. Specifically, { players are assumed to measure their current reward and} to interact in a pairwise fashion, as they exchange information regarding their currently played action and the corresponding reward. On the basis of this communication they possibly revise their strategy by imitating the action played by the other player. {Imitation mechanisms in learning and decision-making processes have been extensively studied in biology, sociology, economics, and marketing~\cite{Traulsen2010,Berg2015}. They have also found engineering applications, e.g., in traffic control problems~\cite{Jiang2014}. Evidence supporting the ubiquity of such mechanisms can be found in empirical studies in human groups \cite{Naber2013}, in the predictive success of imitation-based models on vaccination decisions \cite{Bauch2005}, and on the recent emergence of the role of influencers in social networks, whereby many users rely on their opinion to decide, for instance, which product to buy or which political party to support.}

{There is a substantial literature providing theoretical analysis of imitation dynamics~\cite{Nachbar1990, Hofbauer2000, Sandholm2001,Levine2007}}. In particular, the book~\cite{Sandholm2010} offers an extensive study and review of results of stability and instability for the different kinds of equilibrium points of imitation dynamics. {Most of these results are primarily concerned with local stability and rely on more stringent assumptions on the imitation mechanism than the ones considered in this work. It is only for some specific forms of imitation dynamics and for some classes of games that global stability has been studied. Specifically, the replicator equation has been exhaustively analyzed. While the first results deal with local stability~\cite{Taylor1978, Schuster1983}, conditions for global stability has  have been established for strict stable games~\cite{Hofbauer2009},~\cite[Chapter 7.2]{Sandholm2010}, potential games~\cite[Chapter 7.1]{Sandholm2010}, and matrix games~\cite{Bomze2002,Cressman2014,Riehl2018}. In~\cite{Fox2013}, global stability for the replicator equation for stable games is studied by means of a passivity argument. In~\cite{Ramirez2014}, the replicator dynamics is proposed as a distributed virus mitigation mechanism and asymptotic convergence results are derived for general networks of interactions. Other specific forms of imitation dynamics such as pairwise proportional imitation have been considered, e.g., in~\cite{Hofbauer2009,Barreiro-Gomez2018}. For more general classes of imitation dynamics, global stability results are limited to specific class of games, such as games with strategic substitutes and  strategic complements --- e.g., the best-shot game and the coordination game in~\cite{Cimini2017}, and some public good games~\cite{Govaert2017}.}

{Importantly, most of the studies reviewed above build on the assumption that players of a population interact on a fully mixed structure, where each player interacts with all the other players with the same intensity. However, this assumption is not often realistic, since real-world networks of interactions often have a complex architecture, with clustered populations and different levels of interactions within the same cluster and between different clusters~\cite{McPherson2001,Newman2003}. The results that highlight the role played by the network structure for different evolutionary game dynamics~\cite{Montanari2010, Kreindler2014, Madeo2015} are key to motivate further analysis of imitation dynamics beyond the fully mixed scenario. In~\cite{Barreiro-Gomez2016}, the authors extend some convergence results of learning protocols (including a class of imitation dynamics) to populations in which the interaction pattern is determined by the actions of the players but no {\it a-priori} constraints on the possible interactions between players are considered. Other approaches deal with multi-population games \cite{Hofbauer2009}~\cite[Chapter 2]{Sandholm2010}, which considers a society made by fully mixed populations. Each population has a different reward structure, and the reward is determined by state of the entire society. However, the players that belong to a population do not interact with players from other populations, so their learning mechanism is fully determined by the local interactions and, thus, by the state of the population they belong.}

In this paper, we introduce a novel model of network imitation dynamics assuming that the interaction pattern between the players {of a single population} is governed by a community structure. {This framework ---whereby the players belong to the same population and share the same reward structure,  
but the learning mechanism is dictated by the community structure--- captures many real-world scenarios. For instance, communities can model different age, gender, or social groups, whereby empirical evidence shows that  people tend to establish more interaction within their social group and be more influenced by people of similar age and same gender~\cite{McPherson2001,mossong2008social}.} 

{Mathematically, the model consists of a system of ordinary differential equations (ODEs) coupled by a  community network structure and our first main result consists in a general characterization of its equilibrium points. The result illustrates in general the role of the underlying population game and of the community structure in  determining the set of equilibria. As we shall see, unlike the scenario with a fully mixed population (i.e., single community) in which the equilibrium points are always directly related to the Nash equilibria of the underlying population game, the community network may lead to the emergence of other equilibrium points or enforce some further constraints on the feasible equilibrium points.

The second part of the paper is devoted to the important class of potential population games~\cite{Monderer1996},~\cite[Chapter 3.1]{Sandholm2010}.} Some preliminary results in this direction can be found in~\cite{cdc2017}. Therein, global stability of Nash equilibria has been proved for the fully mixed community-free scenario. {In this paper, we extend that preliminary analysis to a general undirected network of interactions driven by a community structure.} The presence of a non-fully mixed network poses several new technical challenges. Our asymptotic analysis relies on coupling a local stability result that only depends on the assumption that the network is undirected and connected, with a global Lyapunov-LaSalle argument that instead also relies on the assumption that the population game is potential. Our second main result establishes the convergence of the imitation dynamics to {a limit set, characterized in terms of the Nash equilibria of the game and of sub-games obtained by restricting the original game to a subset of actions. When the Nash equilibria of the population game are isolated and fully supported, our result implies convergence to a Nash equilibrium from every fully supported initial population state}. Examples and numerical simulations are offered along with the theoretical results to explain their practical use, offer a better understanding of their implications, and clarify the role of the assumptions made.

In summary, the main contributions of this paper are fourfold: (i) a rigorous formalization of imitation dynamics in population games on community networks; (ii) a characterization of the equilibrium points of these dynamics; (iii) for potential population games over undirected connected networks, a complete analysis of the asymptotic behavior; and (iv) the presentation of several examples to help elucidate the effect of the community network structure.

The rest of the paper is organized as follows. Section~\ref{sec:imitation} introduces population games, community network, and imitation dynamics. In Section~\ref{sec:rest-points}, we characterize the equilibrium points of these systems. In Section \ref{sec:undirected}, we refine our results for undirected networks, while in Section~\ref{sec:potential}, we carry on a complete convergence analysis for potential population games. Section~\ref{sec:conclusion} outlines some future research directions. The Appendix collects the proofs of some technical results.

We end this section by gathering some notational conventions adopted throughout the paper. We denote by $\R$  and $\R_{+}$ the sets of real and nonnegative real numbers, respectively. {For finite sets $\mc A$ and $\mc B$, let $\R^{\mc A}$ (respectively, $\R^{\mc A\times\mc B}$) denote the set of real vectors (matrices) whose entries are indexed by the elements of $\mc A$ ($\mc A\times\mc B$).} The transpose of a vector or matrix $\x$ is denoted as $\x^\top$.  The all-$1$ vector is denoted by $\1$ and ${\sgn}$ denotes the sign function. A directed graph $\mc G=(\mc V,\mc E)$ is the pair of a set of nodes $\mc V$ and a set of directed links $\mc E\subseteq \mc V\times \mc V$: it is said to be connected if for every two nodes $i,j$ in $\mc V$ there exists a directed path from $i$ to $j$.

\section{Description of the model
}\label{sec:imitation}

We consider a continuous of individuals engaged in a single population game. {Each individual chooses their strategy from the same set $\mc A$ and gets a reward that depends exclusively on the chosen strategy and the distribution of strategies within the population. The population is assumed to be structured into communities whose reciprocal interaction is determined by a community network. Individuals update their strategy on the basis of a simple pairwise imitation mechanism confronting their own reward with that of another individual and possibly copying the strategy played by them. The rate at which they establish such pairwise interactions with other individuals is determined by the community network structure (specifically on the communities where the two individuals belong to). Below we present a formal definition of the various concepts.}

{\subsection{The basic ingredients}

{\bf Population game.} Given a finite set of strategies $\mc A$, we denote by }
$$\mc Y:=\left\{\y\in\R_{+}^{\mc A}:\1^\top \y=1\right\}\,$$
{the unitary simplex over $\mc A$. The \emph{population state} is a vector $\y$ in $\mc Y$ whose entries $y_i$ denote the fraction of individuals playing action $i$ in $\mc A$ (also referred to as $i$-players). The reward functions }
$$r_i:\mc Y\to\R\,,\qquad i\in\mc A\,,$$
return the the reward $r_i(\y)$  of any $i$-player as a function of the {population state} $\y$. We refer to a pair $(\mc A, r)$ with the above properties as a \emph{(continuous) population game}.

{{\bf Community network.} Individuals are structured into a finite set $\mc H$ of interacting communities. Given an action $i$ in $\mc A$ and a community $h$ in $\mc H$, $x_{ih}$ denotes the fraction of population residing in $h$ that is playing action $i$. Assembling all these values in a matrix, we obtain $\x$ in $\R_{+}^{\mc A\times\mc H}$ that is called the \emph{system state}. Notice that the population state $\y$ corresponding to a system state $\x$ can be obtained as 
\be\label{y=x1}\y=\x\1\,. \ee
 Communities have fixed relative sizes (possibly not uniform)} described by a {constant} vector $\etav$ in $\mathbb R^{\mc H}$ whose entries $\eta_h>0$ represent the fraction of population belonging to the different communities $h$ in $\mc H$.  {Notice that, as a consequence, the system state $\x$ always verifies the condition \be\label{1x=eta}\1^\top \x=\etav^\top\,.\ee} 
We introduce the set 
$$\mc X:=\left\{\x\in\R_{+}^{\mc A\times\mc H}:\,\eqref{1x=eta}\right\}\,$$
of all admissible {\emph{system states}}. 

The strength of the interactions among individuals in the different communities is described by a {constant nonnegative} matrix $W$ in $\R_{+}^{\mc H\times\mc H}$ with strictly positive diagonal entries.  Specifically, when the system is in state $\x$ in $\mc X$, for every two communities $h$ and $k$ in $\mc H$ and actions $i$ and $j $ in $\mc A$, the product  $$x_{ih}W_{hk}x_{jk}$$ describes the rate at which $i$-players in community $h$ meet $j$-players in community $k$. The triple $\mc G=(\mc H, \etav, W)$ is called a \emph{community network}. To it we canonically associate a directed graph $(\mc H, \mc E)$ with link set $\mc E=\{(h,k)\in\mc H\times \mc H: W_{hk}>0\}$. From now on any graph-theoretic property of $\mc G$ will always be meant as holding true for $(\mc H, \mc E)$. In particular, we will say that $\mc G$ is \emph{connected} if $W$ is irreducible and that $\mc G$ is \emph{undirected} if the matrix $W$ is symmetric, i.e., if $W=W^\top$. 

{{\bf Imitation mechanism.} 
When an individual playing action $i$ meets another individual playing strategy $j$, the former gets informed of the reward $r_j$ that the latter is getting, compares it with their own reward $r_i$, and decides whether to modify their action from $i$ to $j$, thus imitating the other individual. We shall assume that, conditioned on the meeting of the $i$-player with the $j$-player, the former imitates the latter at a rate $f_{ij}(r_i, r_j)$ that depends exclusively on the two current rewards. Since the rewards are functions of the current population state $\y$, from now on we will think such rates as functions $f_{ij}:\mc Y\to \R_{+}$ and, with a slight abuse of notation, write $f_{ij}(\y)$. The functions $f_{ij}$ are then assembled in a matrix-valued function $$\f:\mc Y\to \R_{+}^{\mc A\times \mc A}\,,$$  called the \emph{imitation mechanism} that we assume to be Lipschitz-continuous on its domain $\mc Y$. We want to stress that a $i$-player in order to compute the imitation rate $f_{ij}(\y)$ does not need to know the whole population state $\y$ but just its own reward $r_i(\y)$ and the reward $r_j(\y)$ of a $j$-player that is the only information obtained through their pairwise interaction.}

{\subsection{Network imitation dynamics}
Given a population game $(\mc A,r)$, a community network $\mc G=(\mc H,\etav, W)$, and an imitation mechanism $\f$, 
we consider a dynamical system evolving in continuous time on the space of admissible system states $\mc X$. This is formally defined as }
\be\label{eq:imitation-dynamics}
\dot x_{ih}=\sum_{j\in\mc A}\sum_{k\in\mc H}\big(x_{jh}W_{hk}x_{ik}f_{ji}(\x\1)-x_{ih}W_{hk}x_{jk}f_{ij}(\x\1)\big)\,,
\ee
for every action $i$ in $\mc A$ and community $h$ in $\mc H$. 
{We shall refer to the dynamical system~\eqref{eq:imitation-dynamics} as a \emph{(network) imitation dynamics}. The interpretation is the following. The nonnegative term $$x_{jh}W_{hk}x_{ik}f_{ji}(\x)$$ represents the rate at which $j$-players in community $h$ change their strategy to action $i$ by imitating $i$-players in community $k$. Therefore, summing up over all communities $k$ and actions $j$ we get the total instantaneous increase of the fraction of new $i$-players in community $h$. Similarly, the negative term represents the total instantaneous decrease of the fraction of $i$-players in community $h$ that modify their strategy as a result of the imitation mechanism.

The solutions of the imitation dynamics~(\ref{eq:imitation-dynamics}) satisfy two basic properties. First, the Lipschitz continuity of $\f$ and the fact that $\mc X$ is compact yield global existence and uniqueness of the solution for every initial condition. Second, the support of the solution, that is the subset of actions effectively played, does not change over time. This is formally stated in the following proposition, proved in Appendix \ref{sect:proof-imitation-properties}. 
We first define the \emph{support} of a population state $\y$ in $\mc Y$ and of a system state $\x$ in $\mc X$ as
 $$\mc S_{\y}:=\{i\in\mc A:y_i>0\}\,,\quad \mc S_{\x}:=\mc S_{\x\1}\,,$$
 respectively.}
 
\begin{proposition}\label{proposition:imitations-properties} Consider the dynamical system~(\ref{eq:imitation-dynamics}). Then, for every initial {system state} $\x(0)$ in $\mc X$, 
\begin{enumerate}
\item[(i)] the system~(\ref{eq:imitation-dynamics}) admits a unique solution $(\x(t))_{t\ge0}$ in $\mc X$; 
\item[(ii)] $\mc S_{\x(0)}=\mc S_{\x(t)}$ for every $t\geq 0$\,.
\end{enumerate}
\end{proposition}
For more restrictive conditions on the imitation mechanisms and {single community fully mixed networks} these results were already presented and discussed in~[Chapter 5.4]\cite{Sandholm2010} and~\cite{cdc2017}.  

\begin{figure}
\begin{center}
\subfloat[]{
%
%
\definecolor{mycolor1}{RGB}{66,153,15}

\begin{tikzpicture}

\begin{axis}[%
 axis lines=middle,
 x   axis line style={->},
y   axis line style={->},
ytick style={draw=none},
ymajorticks=false,
xtick style={draw=none},
xmajorticks=false,
width=\ll cm,
height=\hh cm,
at={(0cm,0cm)},
scale only axis,
xmin=-.8,
xmax=1.2,
xlabel={\small $r_j$},
     extra x ticks ={0},
    extra x tick labels={$0$},
ymin=0,
ymax=1,
ylabel={\small $f_{ij}$},
axis background/.style={fill=white},
]
\addplot [color=violet,solid,forget plot, ultra thick]
  table[row sep=crcr]{%
  -.6 	.2\\
.8	.6\\
};

\addplot [color=violet,dashed,forget plot, ultra thick]
  table[row sep=crcr]{%
1	.65714285714\\
.8	.6\\
};

\addplot [color=violet,dashed,forget plot,ultra thick]
  table[row sep=crcr]{%
-.6	.2\\
-.8	.14285714285\\
};

\end{axis}
\end{tikzpicture}%
}\,\subfloat[]{
%
%
\definecolor{mycolor1}{RGB}{66,153,15}

\begin{tikzpicture}

\begin{axis}[%
 axis lines=middle,
 x   axis line style={->},
y   axis line style={->},
ytick style={draw=none},
ymajorticks=false,
xtick style={draw=none},
xmajorticks=false,
width=\ll cm,
height=\hh cm,
at={(0cm,0cm)},
scale only axis,
xmin=-.8,
xmax=1.2,
xlabel={\small $r_j-r_i$},
     extra x ticks ={0},
    extra x tick labels={$0$},
ymin=0,
ymax=1.2,
ylabel={\small $f_{ij}$},
axis background/.style={fill=white},
]
\addplot [color=mycolor1,solid,forget plot, ultra thick]
  table[row sep=crcr]{%
  -.6 	0\\
0	0\\
.8	.8\\
};

\addplot [color=mycolor1,dashed,forget plot, ultra thick]
  table[row sep=crcr]{%
1	1\\
.8	.8\\
};

\addplot [color=mycolor1,dashed,forget plot,ultra thick]
  table[row sep=crcr]{%
-.6	0\\
-.8	0\\
};

\end{axis}
\end{tikzpicture}%
}\,\subfloat[]{
%
%

\definecolor{mycolor2}{RGB}{20,155,204}%
\begin{tikzpicture}

\begin{axis}[%
 axis lines=middle,
 x   axis line style={->},
y   axis line style={->},
ytick style={draw=none},
ymajorticks=false,
xtick style={draw=none},
xmajorticks=false,
width=\ll cm,
height=\hh cm,
at={(0cm,0cm)},
scale only axis,
xmin=-1,
xmax=1,
xlabel={\small $r_j-r_i$},
     extra x ticks ={0},
    extra x tick labels={$0$},
ymin=0,
ymax=1.1,
ylabel={\small $f_{ij}$},
axis background/.style={fill=white}
]
\addplot [color=mycolor2,solid,forget plot,ultra thick]
  table[row sep=crcr]{%
-0.8	0.0831726964939224\\
-0.79	0.0854891394348065\\
-0.78	0.0878639148293013\\
-0.77	0.0902981447029198\\
-0.76	0.092792953117157\\
-0.75	0.0953494648991095\\
-0.74	0.097968804297554\\
-0.73	0.100652093564107\\
-0.72	0.10340045145825\\
-0.71	0.106214991675176\\
-0.7	0.109096821195613\\
-0.69	0.11204703855699\\
-0.68	0.11506673204555\\
-0.67	0.11815697780927\\
-0.66	0.121318837891737\\
-0.65	0.124553358187416\\
-0.64	0.127861566319081\\
-0.63	0.131244469438523\\
-0.62	0.134703051952028\\
-0.61	0.138238273172494\\
-0.6	0.141851064900488\\
-0.59	0.145542328936965\\
-0.58	0.149312934530844\\
-0.57	0.153163715765086\\
-0.56	0.157095468885453\\
-0.55	0.161108949576585\\
-0.54	0.165204870190615\\
-0.53	0.169383896934019\\
-0.52	0.173646647019005\\
-0.51	0.177993685786246\\
-0.5	0.182425523806356\\
-0.49	0.186942613968046\\
-0.48	0.191545348561468\\
-0.47	0.196234056365779\\
-0.46	0.201008999750529\\
-0.45	0.205870371800947\\
-0.44	0.210818293477747\\
-0.43	0.215852810822515\\
-0.42	0.220973892220188\\
-0.41	0.226181425730546\\
-0.4	0.231475216500982\\
-0.39	0.236854984273145\\
-0.38	0.242320360996295\\
-0.37	0.24787088856043\\
-0.36	0.253506016662338\\
-0.35	0.259225100817846\\
-0.34	0.265027400533481\\
-0.33	0.270912077650694\\
-0.32	0.27687819487561\\
-0.31	0.282924714507027\\
-0.3	0.289050497374996\\
-0.29	0.295254302001909\\
-0.28	0.301534783997461\\
-0.27	0.307890495698212\\
-0.26	0.314319886061746\\
-0.25	0.320821300824607\\
-0.24	0.32739298293224\\
-0.23	0.33403307324818\\
-0.22	0.340739611548615\\
-0.21	0.347510537807256\\
-0.2	0.354343693774205\\
-0.19	0.361236824851158\\
-0.18	0.368187582263898\\
-0.17	0.375193525531571\\
-0.16	0.382252125230751\\
-0.15	0.389360766050778\\
-0.14	0.396516750135274\\
-0.13	0.403717300703212\\
-0.12	0.410959565941335\\
-0.11	0.418240623158164\\
-0.1	0.425557483188341\\
-0.09	0.432907095034546\\
-0.08	0.440286350732807\\
-0.07	0.447692090425675\\
-0.0599999999999999	0.45512110762642\\
-0.0499999999999999	0.462570154656251\\
-0.04	0.470035948235428\\
-0.03	0.4775151752082\\
-0.02	0.48500449838059\\
-0.01	0.49250056244938\\
0	0.5\\
0.01	0.50749943755062\\
0.02	0.51499550161941\\
0.03	0.5224848247918\\
0.04	0.529964051764572\\
0.0499999999999999	0.537429845343749\\
0.0599999999999999	0.54487889237358\\
0.07	0.552307909574325\\
0.08	0.559713649267193\\
0.09	0.567092904965454\\
0.1	0.574442516811659\\
0.11	0.581759376841836\\
0.12	0.589040434058665\\
0.13	0.596282699296788\\
0.14	0.603483249864726\\
0.15	0.610639233949222\\
0.16	0.617747874769249\\
0.17	0.624806474468429\\
0.18	0.631812417736102\\
0.19	0.638763175148842\\
0.2	0.645656306225795\\
0.21	0.652489462192744\\
0.22	0.659260388451385\\
0.23	0.66596692675182\\
0.24	0.67260701706776\\
0.25	0.679178699175393\\
0.26	0.685680113938254\\
0.27	0.692109504301788\\
0.28	0.698465216002539\\
0.29	0.704745697998091\\
0.3	0.710949502625004\\
0.31	0.717075285492973\\
0.32	0.72312180512439\\
0.33	0.729087922349307\\
0.34	0.734972599466519\\
0.35	0.740774899182154\\
0.36	0.746493983337662\\
0.37	0.75212911143957\\
0.38	0.757679639003705\\
0.39	0.763145015726855\\
0.4	0.768524783499018\\
0.41	0.773818574269454\\
0.42	0.779026107779812\\
0.43	0.784147189177485\\
0.44	0.789181706522253\\
0.45	0.794129628199053\\
0.46	0.798991000249471\\
0.47	0.803765943634221\\
0.48	0.808454651438533\\
0.49	0.813057386031954\\
0.5	0.817574476193644\\
0.51	0.822006314213753\\
0.52	0.826353352980995\\
0.53	0.830616103065981\\
0.54	0.834795129809385\\
0.55	0.838891050423415\\
0.56	0.842904531114547\\
0.57	0.846836284234914\\
0.58	0.850687065469156\\
0.59	0.854457671063035\\
0.6	0.858148935099512\\
0.61	0.861761726827506\\
0.62	0.865296948047972\\
0.63	0.868755530561477\\
0.64	0.872138433680919\\
0.65	0.875446641812584\\
0.66	0.878681162108263\\
0.67	0.88184302219073\\
0.68	0.88493326795445\\
0.69	0.88795296144301\\
0.7	0.890903178804387\\
0.71	0.893785008324824\\
0.72	0.89659954854175\\
0.73	0.899347906435893\\
0.74	0.902031195702446\\
0.75	0.904650535100891\\
0.76	0.907207046882843\\
0.77	0.90970185529708\\
0.78	0.912136085170699\\
0.79	0.914510860565194\\
0.8	0.916827303506078\\
};

\addplot [color=mycolor2,dashed,forget plot,ultra thick]
  table[row sep=crcr]{%
-1	0.0474258731775668\\
-0.99	0.0487997229988\\
-0.98	0.0502112731902665\\
-0.97	0.0516614354660848\\
-0.96	0.0531511363980637\\
-0.95	0.0546813172159408\\
-0.94	0.0562529335731674\\
-0.93	0.057866955276361\\
-0.92	0.0595243659765015\\
-0.91	0.061226162819908\\
-0.9	0.0629733560569965\\
-0.89	0.0647669686067855\\
-0.88	0.0666080355750907\\
-0.87	0.0684976037243261\\
-0.86	0.0704367308928171\\
-0.85	0.0724264853615177\\
-0.84	0.0744679451660281\\
-0.83	0.0765621973518122\\
-0.82	0.0787103371705352\\
-0.81	0.0809134672154653\\
};

\addplot [color=mycolor2,dashed,forget plot,ultra thick]
  table[row sep=crcr]{%
0.81	0.919086532784535\\
0.82	0.921289662829465\\
0.83	0.923437802648188\\
0.84	0.925532054833972\\
0.85	0.927573514638482\\
0.86	0.929563269107183\\
0.87	0.931502396275674\\
0.88	0.933391964424909\\
0.89	0.935233031393214\\
0.9	0.937026643943004\\
0.91	0.938773837180092\\
0.92	0.940475634023498\\
0.93	0.942133044723639\\
0.94	0.943747066426833\\
0.95	0.945318682784059\\
0.96	0.946848863601936\\
0.97	0.948338564533915\\
0.98	0.949788726809734\\
0.99	0.9512002770012\\
1	0.952574126822433\\
};

\end{axis}
\end{tikzpicture}%
}
\end{center}
\caption{Imitation rates of (a) replicator  (Example~\ref{ex:pairwise}), {(b) pairwise proportional imitation (Example~\ref{ex:pairwise})}, and (c) {sigmoid} imitation (Example~\ref{ex:sigmoid}).}
\label{fig:rates}
\end{figure}
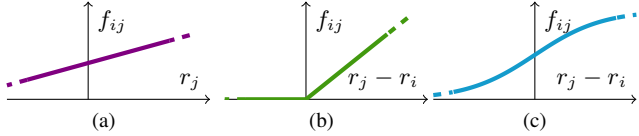

{
In many applications, the imitation rate $f_{ij}$ is nondecreasing function of the difference of rewards $r_j-r_i$. Three examples, which will be used to discuss the theoretical results established in the paper, are presented below. The corresponding imitation rates are plotted in Figure \ref{fig:rates}.}

\begin{example}[Replicator equation]\label{ex:replicator}{ In the special case where  imitation rates are affine functions of the reward, that is, 
\be\label{mech:replicator}f_{ij}(\y)=c+r_j(\y)\ee
for some constant $c>-\min\{r_i(\y):\,\y\in\mc Y,\,i\in\mc A\}$, the imitation dynamics \eqref{eq:imitation-dynamics} reads
\be\label{eq:replicator-network}\dot x_{ih}=\eta_hz_{ih}r_i(\x\1)-x_{ih}\sum_{j\in\mc A}z_{jh}r_j(\x\1)\ee
where $z_{ih}=\sum_{k\in\mc H}x_{ik}W_{hk}$. This dynamics is known as the replicator equation on graphs~\cite{Ohtsuki2006}, and it is extensively used in theoretical biology to model evolutionary dynamics. For a single fully mixed population, i.e., when $|\mc H|=1$, and $W=1$, \eqref{eq:replicator-network} simplifies to 
\be\label{eq:replicator-equation}\dot x_{i}=x_i\left(r_i(\x)-\sum_{j\in\mc A}x_jr_j(\x)\right)\,,\ee
that is the well known replicator equation, see, e.g.,~\cite{Weibull1995,Hofbauer1998,Taylor1978,Schuster1983}.}\end{example}

{\begin{example}[Pairwise proportional imitation]\label{ex:pairwise} Consider the imitation mechanism \be\label{mech:replicator2}f_{ij}(\y)=\max\{r_j(\y)-r_i(\y),0\}\,,\ee
proposed in \cite[Example 4.3.1]{Sandholm2010}. For a single fully mixed population, \eqref{mech:replicator2} again simplifies to the replicator equation \eqref{eq:replicator-equation}. 
\end{example}}

\begin{example}[{Sigmoid imitation}]\label{ex:sigmoid}
Let 
\begin{equation}\label{eq:logistic}f_{ij}(\y)=\frac{1}{1+\exp\left\{-K_{ij}\big(r_j(\y)-r_i(\y)\big)\right\}},\end{equation}
where $K_{ij}>0$ are constants possibly different for each pair of actions $(i,j)$ in $\mc A\times\mc A$. 
{This is the logistic function often used in the literature to model learning curves and adoption of innovation~\cite{Bass1969}. Other sigmoid functions (such as the hyperbolic tangent or the arctangent) also fit in this framework.}
\end{example}

{The theory developed in this paper encompasses the cases in which the imitation rates are functions of the rewards (as in the replicator equation) and of rewards' difference as in the last two examples. The somewhat minimal assumption needed for our results is reported below. It states that, if an action $j$ gives a greater reward than another action $i$, then the imitation rate from $i$ to $j$ is greater than the one from $j$ to $i$.}
\begin{assumption}\label{assumption} For every two actions $i,j$ in $\mc A$ and population state  $\y$ in $\mc Y$
\be\label{assumption:fij}{\sgn}\big(f_{ij}(\y)-f_{ji}(\y)\big)={\sgn}\big(r_j(\y)-r_i(\y)\big).\ee
\end{assumption}
\begin{remark}
The class of imitation mechanisms satisfying Assumption \ref{assumption} includes and is broader than the ones typically considered in the literature~\cite[Chapter 5.4]{Sandholm2010}, which satisfy the stricter assumption
\be\label{assumption:sandholm}
r_i(\y)\geq r_j(\y)\iff f_{\ell i}(\y)-f_{i\ell}(\y)\geq f_{\ell j}(\y)-f_{j\ell}(\y)\,,
\ee
for every two actions $i,j,\ell$ in $\mc A$ and $\y$ in $\mc Y$.  {We notice that while Examples \ref{ex:replicator} and \ref{ex:pairwise} satisfy~\eqref{assumption:sandholm}, in general Example~\ref{ex:sigmoid} does not.}
\end{remark}

For some {---but not all---} of the results presented in this paper, we will need an {additional assumption on the imitation mechanism} ensuring that, when many actions give the same reward, then the imitation rates between these actions are always nonzero and individuals have no clear preference for one of them. We formalize these ideas in the following.
\begin{assumption}\label{assumption2} For every three distinct actions $i, j,\ell$ in $\mc A$ and  population state $\y$ in ${\mc Y}$
$$r_i(\y)=r_{j}(\y)=r_{\ell}(\y)\;\Rightarrow\; f_{ij}(\y)=f_{i\ell}(\y)>0.$$
\end{assumption}

{While all the examples verifies Assumption~\ref{assumption}, Assumption~\ref{assumption2} holds true in Examples \ref{ex:replicator} and \ref{ex:sigmoid}, but not in Example \ref{ex:pairwise}.}

\section{Equilibrium  points of the imitation dynamics}\label{sec:rest-points}

{
This section is entirely devoted to the study of the set $\mc Z\subseteq\mc X$ of equilibrium points of the network imitation dynamics~\eqref{eq:imitation-dynamics}. Such equilibrium points may  be intuitively expected to relate to the Nash equilibria of the underlying population game. The set of such Nash equilibria will be denoted as 
$$\nash:=\left\{\y\in\mc Y:\,y_i>0\Rightarrow r_i(\y)=\max_{j\in\mc A}r_j(\y)\right\}\,.$$
In a Nash equilibrium $\y$ in $\nash$, all actions played by a nonzero fraction of players give the same reward, i.e., $r_i(\y)=r_j(\y)$, for every $i,j$ in $\mc S_{\y}$, and such reward is not smaller than the reward of all remaining actions, i.e., $r_i(\y)\ge r_j(\y)$, for every $i$ in $\mc S_{\y}$ and $j$ in $\mc A\setminus\mc S_{\y}$. Since point (ii) of Proposition \ref{proposition:imitations-properties} ensures that actions that are not played at a certain time will remain not played at any future time and thus will never play any role in the dynamics, another natural set of population states to consider is the set of \emph{restricted Nash equilibria}
$$\rnash:=\left\{\y\in\mc Y:\,y_i>0,\,y_j>0\Rightarrow r_i(\y)=r_j(\y)\right\}\,.$$
In a restricted Nash equilibrium $\y$ in $\rnash$, all actions played by a nonzero fraction of players give the same reward, however, unless $\y$ is also a Nash equilibrium, there exist some actions not played by anyone that give a strictly higher reward. Notice that every restricted Nash equilibrium $\y$ in $\rnash$ can be interpreted as a Nash equilibrium of the sub-game obtained by restricting the action set to $\mc S_{\y}$. We also introduce the sets of system states
$$\xnash:=\{\x\in\mc X\,:\, \x\1\in\nash\},\quad \xrnash:=\{\x\in\mc X\,:\, \x\1\in\rnash\}\,,$$
associated to Nash and restricted Nash equilibria, respectively.

For the single community, fully mixed population case, it was proven in \cite{cdc2017} that the set of equilibrium points of the imitation dynamics coincides with the set of restricted Nash equilibria, i.e., in this case $\mc Z=\xrnash=\rnash$.  
As we will see,  in general this result cannot be extended as such to the case of a nontrivial community network, since two types of issues arise. Perhaps not surprisingly, a first issue is related to connectivity and the lack thereof: if the community network is not connected, equilibrium points can exhibit actions with different rewards played in different connected components. A second issue is perhaps less obvious: in certain cases, the community network structure imposes extra constraints on the equilibrium points, namely that the action's distribution is the same on every single community. This is captured by the following definition. We say that a system state $\x$ in $\mc X$ is \emph{balanced} if, with $\y=\x\1$, we have  that
\begin{equation}\label{eq:balanced}\x_{ih}=y_i\eta_h\,,\qquad\forall\,i\in\mc A\,,\quad\forall\,h\in\mc H\,.\end{equation}
This is equivalent to write that $\x=\x\1{\etav^\top}$.
We denote by $\xnashbal$ and $\xrnashbal$ the subsets of the balanced system states contained, respectively, in $\xnash$ and in $\xrnash$.

The following states the main results of this section.}
\begin{theorem}\label{teo:restpoints} 
Consider a population game  $(\mc A,r)$, a connected community network $\mc G={(\mc H, \etav, W)}$, and an imitation mechanism $\f$ satisfying Assumption \ref{assumption}. 
{Then, \be\label{fund-inclusion}\xrnashbal\subseteq \mc Z\subseteq \xrnash\,.\ee Moreover,
\begin{enumerate}
\item[(i)] if $f_{ij}(\y)> 0$ for every $\y$ in $\mc Y$ and $i,j$ in $\mc A$,  then $$\mc Z=\xrnashbal\,;$$
\item[(ii)] if $f_{ij}(\y)= 0$ for every$\y$ in $\mc Y$ and $i,j$ in $\mc A$ such that $r_i(\y)=r_j(\y)$,  then $$\mc Z=\xrnash\,.$$
\end{enumerate}}
\end{theorem}

\begin{IEEEproof}
{We first show that ${\xrnashbal}\subseteq \mc Z$. Let $\x$ in $\xrnashbal$ and put $\y=\x\1$.}  If $i$ does not belong to $\mc S_{\x}$, then $x_{ih}=0$ for every $h$ in $\mathcal H$ and this implies that, for such $i$, the right-hand side of~\eqref{eq:imitation-dynamics} is $0$. Suppose now that $i$ in $\mc S_{\x}$ and $h$ in $\mc H$. For any $j$ in $\mc S_{\x}$, the fact that $r_i(\y)=r_j(\y)$ and Assumption~\ref{assumption} yield that $f_{ij}(\y)=f_{ji}(\y)$. Using this and \eqref{eq:balanced}, we obtain from~\eqref{eq:imitation-dynamics} that
$$\dot x_{ih}=\sum_{k\in\mc H}W_{hk}\sum_{j\in\mc S_{\x}} [y_j\eta_hf_{ij}(\x)y_i\eta_k-y_i\eta_h f_{ij}(\x)y_j\eta_k]=0,$$
which yields the claim. 

We now show that  ${\mc Z\subseteq \xrnash}$. For $\x$ in $\mc Z$ and $\y=\x\1$,  define
$$\mc S^*_\x:=\displaystyle\argmin_{j\in\mc S_\x}r_j(\y)\,,\qquad  \mc S^{**}_\x:=\displaystyle\argmax_{j\in\mc S_\x}r_j(\y)$$
We will show that $\mc S^*_\x\subseteq \mc S^{**}_\x$ (indeed this implies that $r_i(\y)$ is constant over $S_\x$ and thus $\x$ in $\xrnash$).
We define
$$\ba{l}
\mc H^{**}:=\{h\in\mc H:\exists\,i\in\mc S_\x^{**},\; x_{ih}>0 \}\,.\ea$$
First, we show that $\mc H^{**}=\mc H$, that is, in the system state $\x$, in all communities there are players that play an action achieving the maximum reward. Indeed, if by contradiction $\mc H^{**}$ were a proper subset of $\mc H$, since $\mc G$ is connected, we would find $h$ in $\mc H\setminus \mc H^{**}$ and $h'$ in $\mc H^{**}$ such that $W_{hh'}>0$. Let $i$ in $\mc S^{**}_\x$ be such that $x_{ih'}>0$ and $i'$ in $\mc S_\x$ be such that $x_{i'h}>0$. Note that $x_{ih}=0$ by the way $h$ has been chosen 
and $f_{i'i}(\y)>0$ because of Assumption~\ref{assumption}. From~\eqref{eq:imitation-dynamics}, we then obtain
$$\dot x_{ih}=\sum\limits_{k\in\mc H}W_{hk}\sum\limits_{j\in\mc A} f_{ji}(\y)x_{jh}x_{ik},\\
\geq W_{hh'}f_{i'i}(\y)x_{i'h}x_{ih'}>0.
$$
which contradicts the fact that $\x$ is an equilibrium point.

{Second, we fix any $i$ in $\mc S^*_\x$ and we assume, by contradiction, that $i\not\in\mc S^{**}_\x$.}  Since $\x$ is an equilibrium point, then, for every community $h$ in $\mc H$, the right-hand side of~\eqref{eq:imitation-dynamics} equals  $0$, so that
\be\label{estim-equilibrium1}
\begin{array}{rcl}0&=&\displaystyle\sum\limits_{k\in\mc H}W_{hk}\sum\limits_{j\in\mc A} [f_{ji}(\y)x_{jh}x_{ik}-f_{ij}(\y)x_{ih}x_{jk}]\\
&=&\displaystyle\sum\limits_{k\in\mc H}W_{hk}\sum\limits_{j\in\mc A} [f_{ji}(\y)-f_{ij}(\y)]x_{jh}x_{ik}\\
&&\displaystyle+\sum\limits_{k\in\mc H}W_{hk}\sum\limits_{j\in\mc A} f_{ij}(\y)[x_{jh}x_{ik}-x_{ih}x_{jk}]\,.
\end{array}
\ee
Focusing on the first term of the final expression in~\eqref{estim-equilibrium1}, notice that, since $i$ is in $\mathcal S_\x^*$, then $r_i(\y)\leq r_j(\y)$ for all $j$ in $\mathcal S_\x$. Hence, Assumption~\ref{assumption} implies that $f_{ji}(\y)-f_{ij}(\y)\leq 0$, yielding 
\be\label{estim-equilibrium2}\sum\limits_{k\in\mc H}W_{hk}\sum\limits_{j\in\mc A} [f_{ji}(\y)-f_{ij}(\y)]x_{jh}x_{ik}\leq 0\,.\ee
Now, rewrite the second term of the final expression in~\eqref{estim-equilibrium1} as
\be\label{estim-equilibrium3}\begin{array}{l}
\displaystyle\sum\limits_{k\in\mc H}W_{hk}\sum\limits_{j\in\mc A} f_{ij}(\y)[x_{jh}x_{ik}-x_{ih}x_{jk}]\\
\displaystyle\qquad=z_h\sum\limits_{k\in\mc H}W_{hk}x_{ik}-x_{ih}\sum\limits_{k\in\mc H}W_{hk}z_k\,,
\end{array}\ee
where \be\label{zh}z_h=\sum_{j\in\mc A}f_{ij}(\y)x_{jh}\,.\ee
From (\ref{estim-equilibrium1}),~(\ref{estim-equilibrium2}), and~(\ref{estim-equilibrium3}), we obtain that
\be\label{estim-equilibrium4}
z_h\sum\limits_{k\in\mc H}W_{hk}x_{ik}-x_{ih}\sum\limits_{k\in\mc H}W_{hk}z_k\geq 0\,.
\ee
Since $\mc H^{**}=\mc H$ and since we have assumed that $i\not\in\mc S^{**}_\x$, we have that $z_h>0$ for every $h$ in $\mc H$. Moving the second term of~\eqref{estim-equilibrium4} to the right-hand side and dividing both sides by the strictly positive quantity $z_h\sum_{k\in\mc H}W_{hk}z_k$, we then obtain the inequality
\be\label{estim-equilibrium45}
\sum\limits_{k\in\mc H}Q_{hk}\alpha_k\geq \alpha_h\,,
\ee
where
\be\label{estim-equilibrium5}
Q_{hk}=\frac{W_{hk}z_k}{\sum_{\ell\in\mc H}W_{h\ell}z_\ell},\quad\text{and}\quad \alpha_k=\frac{x_{ik}}{z_k}\,.
\ee
Since $Q$ is an irreducible stochastic matrix, standard matrix theory allows to deduce that $\alpha$ is a constant vector and that, consequently, ~\eqref{estim-equilibrium45} is satisfied with equality for every $h$ in $\mc H$. This implies that also~\eqref{estim-equilibrium4} is satisfied with equality. Since the sum of the two terms on the right-hand side of~\eqref{estim-equilibrium1} is equal to $0$ and the second one is $0$, it follows that also~\eqref{estim-equilibrium2} is satisfied with equality. {Since $f_{ji}(\y)-f_{ij}(\y)\leq 0$ for all $j$ in $\mc A$, we obtain that 
\be\label{estim-equilibrium2bis}
[f_{ji}(\y)-f_{ij}(\y)]\sum\limits_{k\in\mc H}W_{hk}x_{jh}x_{ik}= 0.
\ee
for every $h$ in $\mc H$ and $j$ in $\mc A$. }
For every $h$ in $\mc H$ such that $x_{ih}>0$, let $j$ in $\mc S^{**}_x$  be such that $x_{jh}>0$ (there exists one since  $\mc H^{**}=\mc H$). From~\eqref{estim-equilibrium2bis},  we have 
$f_{ji}(\y)=f_{ij}(\y)$, so that $r_i(\y)=r_j(\y)$ and then $i$ belongs to $\mc S^{**}_x$. This is a contradiction. Therefore, $\mc S^*_\x=\mc S^{**}_\x$, equivalently,  $\x$ in $\xrnash$.

\begin{figure}
\begin{center}
\subfloat[]{	
\begin {tikzpicture}
\node[draw=red, fill=red!10, ultra thick,circle] (1) at (0,0) {$a$};
\node[draw=blue, fill=blue!10, ultra thick, circle] (2) at (1.2,.8) {$b$};
\node at (1.2,-.4) {$$};
\path[->, >=latex] (1) edge  [loop right] node {$W_{aa}$} (1);
\path[->, >=latex] (2) edge  [loop left] node {$W_{bb}$} (2);
\label{fig:topI}
\end{tikzpicture}}\,\,
\subfloat[]{	
\begin {tikzpicture}
\node[draw=red, fill=red!10, ultra thick,circle] (1) at (0,0) {$a$};
\node[draw=blue, fill=blue!10, ultra thick, circle] (2) at (1.2,.8) {$b$};
\draw[->, >=latex] (2) edge node[below=4pt, right=1pt] {{$W_{ba}$}}(1) ;
\path[->, >=latex] (1) edge  [loop left] node[below] {$W_{aa}$} (1);
\path[->, >=latex] (2) edge  [loop left] node[left] {$W_{bb}$} (2);
\label{fig:topD}
\end{tikzpicture}}\,\,
\subfloat[]{	
\begin {tikzpicture}
\node[draw=red, fill=red!10, ultra thick,circle] (1) at (0,0) {$a$};
\node[draw=blue, fill=blue!10, ultra thick, circle] (2) at (1.2,.8) {$b$};
\draw[->, >=latex] (1) edge[bend left] node[above=4pt] {{$W_{ab}$}}(2) ;
\draw[->, >=latex] (2) edge[bend left] node[below=4pt] {{$W_{ba}$}}(1) ;
\path[->, >=latex] (1) edge  [loop left] node[below] {$W_{aa}$} (1);
\path[->, >=latex] (2) edge  [loop right] node[below=4pt] {$W_{bb}$} (2);
\label{fig:topC}
\end{tikzpicture}}
\end{center}
\caption{Community networks $\mc G=(\mc H,\etav, W)$ analyzed in the examples.}
\label{fig:top}
\end{figure}
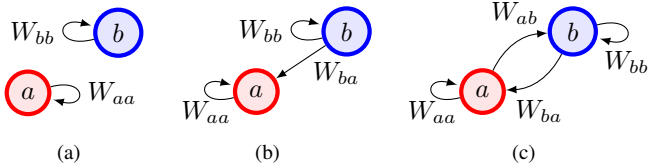

{\bf (i)} {By~\eqref{fund-inclusion}, we only need to prove the inclusion $\mc Z\subseteq\xrnashbal$.} Fix $\x$ in $\mc Z$ and put $\y=\x\1$. Since, by~\eqref{fund-inclusion}, $\x$ in $\xrnash$, $r_i(\y)$ is constant for all $i$ in $\mc S_\x$. We fix any $i$ in $\mc S_x$ and any $h$ in $\mc H$ and we use some of the algebraic computation developed in the above. 
From \eqref{estim-equilibrium1} and the fact that, in this case, \eqref{estim-equilibrium2} trivially holds true as an equality, we obtain that also \eqref{estim-equilibrium4} holds true as an equality. Notice that, since $f_{ij}(\y)> 0$ for every $\y$ in $\mc Y$ and $j$ in $\mc A$, it follows that $z_k>0$ (see \eqref{zh}) for every $k$ in $\mc H$. We can thus consider \eqref{estim-equilibrium45} that, consequently, also holds true as an equality. Since the vector $\mb\alpha$ with entries as  in~\eqref{estim-equilibrium5} must be constant, there exist constants $c_i$, for $i$ in $\mc S_x$, such that   
\be\label{estim-equilibrium6}
x_{ih}=c_i\sum_{j\in\mc S_\x}f_{ij}(\y)x_{jh}\,,
\ee
for $i$ in $\mc S_\x$ and $h$ in $\mc H$. 
Summing both sides of~\eqref{estim-equilibrium6} over all $h$ in $\mc H$, we obtain the following equality, for every $i$ in $\mc S_\x$,
\be\label{estim-equilibrium7}
y_i=c_i\sum_{j\in\mc S_\x}f_{ij}(\y)y_j.
\ee
Equations~(\ref{estim-equilibrium6}) and~(\ref{estim-equilibrium7}) yield
$$y_i^{-1}x_{ih}=\sum_{j\in\mc S_\x}F_{ij}y_j^{-1}x_{jh},$$
where
$$F_{ij}=\frac{f_{ij}(\y)y_j}{\sum_{j'\in\mc S_\x}f_{ij'}(\y)y_{j'}}.$$
Since the matrix $F$ is stochastic and with positive entries, it follows that, for $h$ in $\mc H$, there exists a constant $\rho_h$ such that 
$y_i^{-1}x_{ih}=\rho_h$, for all $i$ in $\mc S_\x$. Multiplying by $y_i$ and summing over $i$ in $\mc S_\x$ we obtain that $\rho_h=\eta_h$. This proves that $\x$ belongs to $\mc Z_{\etav}$.

{\bf (ii)} {By (\ref{fund-inclusion}), we only need to prove the inclusion $\xrnash\subseteq \mc Z$. Fix $\x$ in $\xrnash$. If $i$ does not belong to $\mc S_\x$, then $x_{ih}=0$ for every $h$ in $\mc H$ and this implies that, for such value of $i$, the right-hand side of~\eqref{eq:imitation-dynamics} is $0$.
If $i$ in $\mc S_\x$, we compute the right hand side as follows:
$$\ba{l}\sum\limits_{j\in\mc A}\sum\limits_{k\in\mc H}\big(x_{jh}W_{hk}x_{ik}f_{ji}(\x\1)-x_{ih}W_{hk}x_{jk}f_{ij}(\x\1)\big)\\=
\sum\limits_{j\in\mc S_\x}\sum\limits_{k\in\mc H}\big(x_{jh}W_{hk}x_{ik}f_{ji}(\x\1)-x_{ih}W_{hk}x_{jk}f_{ij}(\x\1)\big)=0\ea$$
where the last equality, which completes the proof, follows from the fact that when $i,j\in S_\x$, we have that $r_i(\x\1)=r_j(\x\1)$, hence $f_{ij}(\x\1)=f_{ji}(\x\1)=0$.}
\end{IEEEproof}\medskip

{Notice that, in the proof of the first inclusion in \eqref{fund-inclusion}, i.e., $\xrnashbal\subseteq \mc Z$, we have not made any use of the connectivity assumption. Hence, this inclusion holds true for every community graph. 
Instead the proof of the inclusion $\mc Z\subseteq \xrnash$ relies on the connectivity assumption. In the following, we present two examples illustrating how, in the absence of connectivity, equilibrium points not corresponding to restricted Nash equilibria can indeed show up.}

\begin{figure}
\centering
\begin{tikzpicture}
\definecolor{mycolor1}{rgb}{0.00000,0.44700,0.74100}%
\definecolor{mycolor2}{rgb}{0.85000,0.32500,0.09800}%
\definecolor{mycolor3}{rgb}{1.00000,1.00000,0.00000}%
\definecolor{mycolor4}{rgb}{0.92900,0.69400,0.12500}%
\definecolor{mycolor5}{rgb}{0.49400,0.18400,0.55600}%
\definecolor{mycolor6}{rgb}{0.46600,0.67400,0.18800}%
\definecolor{mycolor7}{rgb}{0.30100,0.74500,0.93300}%
\definecolor{mycolor8}{rgb}{0.63500,0.07800,0.18400}%
\begin{axis}[
 x   axis line style={->},
y   axis line style={->},
at={(0,0)},
    width=\l cm,
height=\h cm,
scale only axis,
xmin=0,
xmax=0.7,
xlabel={$x_{1a}$},
 extra y ticks ={0},
    extra y tick labels={$0$},
     extra x ticks ={0},
    extra x tick labels={$0$},
ylabel style={yshift=-.5cm}, 
ymin=0,
ymax=.3,
ylabel={$x_{1b}$},
axis background/.style={fill=white},
]

\addplot [color=mycolor1,ultra thick]
  table[row sep=crcr]{%
0	0.1\\
0	0.107763627273481\\
0	0.125928404947781\\
0	0.142486114271349\\
0	0.15562724711746\\
0	0.15882034928677\\
0	0.159771153777155\\
0	0.159956401062634\\
0	0.159991723389394\\
0	0.15999842988446\\
0	0.159999702179782\\
0	0.159999943510723\\
0	0.159999984306645\\
};

\addplot [color=black, draw=none, mark size=4pt,  mark=x]
  table[row sep=crcr]{%
0	0.16\\
};

\addplot [color=mycolor1, draw=none, mark=*, mark options={solid, fill=mycolor1, mycolor1}]
  table[row sep=crcr]{%
0	0.1\\
};

\addplot [color=mycolor4,ultra thick]
  table[row sep=crcr]{%
0	0.25\\
0	0.232632359506173\\
0	0.213576766374573\\
0	0.197978097976907\\
0	0.185500182411992\\
0	0.175855101319358\\
0	0.168773484498718\\
0	0.163958549450008\\
0	0.160972014885516\\
0	0.160180873112255\\
0	0.160034186509366\\
0	0.160006480011474\\
0	0.160001228939423\\
0	0.160000233093115\\
0	0.160000044211659\\
0	0.160000007555716\\
};

\addplot [color=mycolor4, draw=none, mark=*, mark options={solid, fill=mycolor4, mycolor4}]
  table[row sep=crcr]{%
0	0.25\\
};

\addplot [color=mycolor6,ultra thick]
  table[row sep=crcr]{%
0.6	0.2\\
0.621698232500224	0.215203505259588\\
0.667050224652458	0.252979840703003\\
0.693383946347378	0.282633809777203\\
0.698787847858323	0.294277285801736\\
0.699987788259417	0.298879670216688\\
0.699999897478035	0.299804754510605\\
0.700000311530585	0.300293450604447\\
0.699999072907841	0.299567130511178\\
0.700001681008585	0.300369124819232\\
0.699997530534822	0.299751625578151\\
0.700004989064264	0.300238389061055\\
0.699986117816265	0.299676228231424\\
0.700034362364771	0.30038140770461\\
0.699936536754706	0.299682184820344\\
0.700116544480662	0.300244927112441\\
0.699699189557759	0.299766789578344\\
0.700464708729572	0.300041177990236\\
0.699278552814223	0.300164226367683\\
0.701010175649036	0.299618365217206\\
0.699163488068218	0.300375854377866\\
0.700525890811319	0.299744561633576\\
0.699530915846344	0.300230971450234\\
0.700576943171477	0.29971250468961\\
};

\addplot [color=mycolor6, draw=none, mark=*, mark options={solid, fill=mycolor6, mycolor6}]
  table[row sep=crcr]{%
0.6	0.2\\
};

\addplot [color=mycolor8,ultra thick]
  table[row sep=crcr]{%
0.4	0.3\\
0.431488805903474	0.29575967849252\\
0.510988897352516	0.288777141275756\\
0.585141019925595	0.286683622320858\\
0.651503102990319	0.289027816600877\\
0.68245395820868	0.293160470712229\\
0.693951637990825	0.296322170355447\\
0.700645810428437	0.299028036763562\\
0.700008265482474	0.299880617845664\\
0.699985153616329	0.300105009335094\\
0.700098587526812	0.299618265973108\\
0.699910105094227	0.300167434524561\\
0.700081868099045	0.299913614002013\\
0.699624079912911	0.300293250219915\\
0.700266691973089	0.299838835319284\\
0.699809975298631	0.300102011670717\\
0.70058993753908	0.299693747691778\\
0.699633724230045	0.300184627631478\\
0.700225903362897	0.299886419226892\\
0.69943923579406	0.300280641698527\\
0.700397497746452	0.299800345193337\\
0.699716403415489	0.300141610474763\\
0.7005933614443	0.299702864992178\\
0.699561780616849	0.300218265906694\\
0.700321407518133	0.299839001024902\\
0.699409235570111	0.300294928950669\\
0.700007356244814	0.299995960274391\\
};

\addplot [color=mycolor8, draw=none, mark=*, mark options={solid, fill=mycolor8, mycolor8}]
  table[row sep=crcr]{%
0.4	0.3\\
};

\addplot [color=mycolor2,ultra thick]
  table[row sep=crcr]{%
0.1	0.2\\
0.1082695573237	0.200043111781111\\
0.158064063608147	0.201627461932259\\
0.252417577401345	0.208173479429188\\
0.352908673849635	0.218337364029995\\
0.434919523243021	0.228911366828489\\
0.512381471601898	0.241255899269507\\
0.586685828734432	0.256301859663132\\
0.661064392573683	0.276706563311806\\
0.689665778159579	0.28961176881416\\
0.698088162378164	0.296267577367675\\
0.700302052501414	0.299535361560631\\
0.699817402584741	0.300152846805949\\
0.700277451472736	0.299817880503336\\
0.699229238413244	0.300443610414454\\
0.700405916837863	0.29978655299179\\
0.69978439645622	0.300109085672441\\
0.700494968904871	0.299750666355447\\
0.699598700200762	0.30020049639002\\
0.700323433639996	0.299837852469024\\
0.699367324108361	0.300315966567943\\
0.701029968293256	0.29948289192512\\
0.699248470876798	0.300373446045891\\
0.700421021092586	0.299788650842432\\
0.699561064795934	0.300219168886221\\
0.700705725655586	0.299646219855322\\
0.699137435099621	0.300429580092631\\
0.700628958084606	0.29968395427093\\
0.699533116000281	0.300232778741122\\
0.70053008394358	0.299734391838135\\
0.700193302600173	0.299903000400886\\
};

\addplot [color=mycolor2, draw=none, mark=*, mark options={solid, fill=mycolor2, mycolor2}]
  table[row sep=crcr]{%
0.1	0.2\\
};

\addplot [color=mycolor5,ultra thick]
  table[row sep=crcr]{%
0.5	0.05\\
0.506545490296779	0.0540181047437406\\
0.537217978982053	0.07453263326414\\
0.57778401758122	0.106876598730283\\
0.614204602861813	0.142933810207984\\
0.645407320786537	0.18184927522035\\
0.671362704156556	0.223508457351133\\
0.691654429447317	0.266465749267926\\
0.698379104936525	0.288576372217139\\
0.699715795285238	0.296480823518234\\
0.699950502287598	0.298950012032489\\
0.700014489775373	0.300010535804319\\
0.699995755845236	0.300001794267509\\
0.700077193718042	0.299964522661959\\
0.69972576166611	0.300131491651809\\
0.700418428911788	0.29979433826212\\
0.699358721542945	0.300317164428264\\
0.70085885546574	0.299570685295634\\
0.699307255011399	0.300344153832504\\
0.700484817478817	0.299756894403653\\
0.699497291157078	0.300250778551245\\
0.700675710903944	0.299661261278402\\
0.699241102836319	0.300378027526093\\
0.700619961170942	0.299688741892044\\
0.699480758570349	0.300258858878686\\
0.700563637526316	0.299717479429762\\
0.699319806646214	0.300339085153837\\
0.700688277857501	0.299654597034173\\
0.699445026334461	0.300276473018328\\
};

\addplot [color=mycolor5, draw=none, mark=*, mark options={solid, fill=mycolor5, mycolor5}]
  table[row sep=crcr]{%
0.5	0.05\\
};

\addplot [color=black, draw=none, mark=*]
  table[row sep=crcr]{%
0	0\\
};
\addplot [color=gray, draw=none, mark=*]
  table[row sep=crcr]{%
0	0.3\\
};

\addplot [color=black, draw=none, mark=*]
  table[row sep=crcr]{%
0.7	0.3\\
};

\end{axis}

\end{tikzpicture}
\caption{Orbits of the system state for the imitation dynamics presented in Example~\ref{ex:dir} with different initial conditions (colored circles). Parameters: $\etav=(0.7,0.3)$, $W_{aa}=W_{bb}=1$, $W_{ba}=0.2$ and $W_{ab}=0$. Black circles are states in $\mc X^\bullet$ and crosses are equilibrium points not in $\mc Z$.}\label{fig:ex3}
\end{figure}
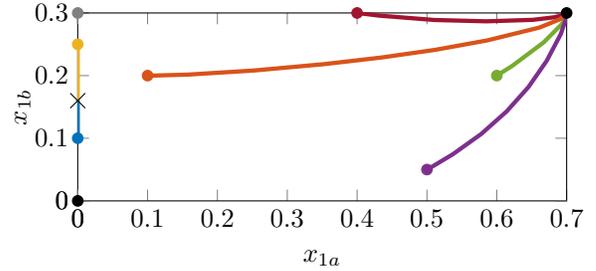

\begin{example}\label{ex:dir}
Consider a population game $(\mc A, r)$ with binary action set $\mc A=\{0,1\}$ and constant reward functions $r_0(\y)=0$ and $r_1(\y)=1$. {The set $\xrnash$ consists of two point: the Nash equilibrium in which the entire population plays action $1$ and the restricted Nash equilibria in which the entire population plays $0$.} Let the imitation rates be constant and given by $f_{01}(\y)=2$ and $f_{10}(\y)=1$. {Consider two possible community networks $\mc G_i=(\mc H, \etav, W^i)$ for $i=1,2$, both consisting of two communities $\mc H=\{a,b\}$, any $\etav$, and weight matrix $W^{(i)}$ such that 
$$W^{(1)}_{ab}=W^{(1)}_{ba}=0,\quad\text{and}\quad W^{(2)}_{ab}=0,\; W^{(2)}_{ba}>0\,.$$

For the community network $\mc G_1$, that is displayed in Figure \ref{fig:topI}, the two communities $a$ and $b$ are isolated: as a consequence, the system state $\x$ in $\mc X$ such that $x_{0a}=\eta_a$ and $x_{1b}=\eta_b$ is an equilibrium point of the network imitation dynamics \eqref{eq:imitation-dynamics}, but it is not in $\xrnash$.

On the other hand,} for the community network $\mc G_2$, that is displayed in Figure~\ref{fig:topD}, since $x_{0h}+x_{1h}=\eta_h$, for $h=a,b$, the imitation dynamics~\eqref{eq:imitation-dynamics} reduce to the planar system
\be\label{eq:imit3I}\left\{\ba{lll}
\dot x_{1a}&=&\ds W^{(2)}_{aa}x_{1a}(\eta_a-x_{1a})\\
\dot x_{1b}&=&\ds W^{(2)}_{bb}x_{1b}(\eta_b-x_{1b})+2W^{(2)}_{ba}(\eta_b-x_{1b})x_{1a}\\&&-W^{(2)}_{ba}x_{1b}(\eta_a-x_{1a}).
\ea\right.
\ee
One can verify that, if $W^{(2)}_{ba}\eta_a<\eta_bW^{(2)}_{bb}$, then the system state
$$x=\left(\begin{matrix}\eta_a&W^{(2)}_{ba}\eta_a/W^{(2)}_{bb}\\0&\eta_b-W^{(2)}_{ba}\eta_a/W^{(2)}_{bb}\end{matrix}\right)$$
is an equilibrium point of the network imitation dynamics~\eqref{eq:imit3I}. Even in this case, this system state does not belong to $\xrnash$ and we notice that not even at the level of a single community the action distribution is necessarily a Nash equilibrium.
{ Observe that trajectories with initial condition $x_{1a}=0$ converge to this equilibrium point, as shown} in Figure \ref{fig:ex3}.
\end{example}

\section{Imitation dynamics on undirected community networks and for potential population games}
{Here, we study more in detail the imitation dynamics \eqref{eq:imitation-dynamics} in the special case when the community network is undirected and connected and the population game is potential, establishing a global convergence result. Specifically, in this scenario, when both Assumptions 1 and 2 hold true and under some assumptions on the set of Nash equilibria of the game, the solution $\x(t)$ of \eqref{eq:imitation-dynamics} converges to an equilibrium point in $\xnashbal$, namely a balanced system state whose corresponding population state is a Nash equilibrium.

First, we present some additional properties of the imitation dynamics when the community network is undirected and connected; then, we introduce potential population games, focusing our analysis on this class of games.}

\subsection{Imitation dynamics on undirected community networks} \label{sec:undirected}

{If the community network is undirected and connected, the imitation dynamics in \eqref{eq:imitation-dynamics} satisfies some additional properties. The first one has an interest of its own and asserts that the set of system states $\xrnash$ is always invariant for the imitation dynamics, even when not all the states in $\xrnash$ are equilibrium points of \eqref{eq:imitation-dynamics} (see Theorem \ref{teo:restpoints}). Precisely, we show that if the imitation dynamics starts  with an initial system state $\x(0)$ in $\xrnash$, then  \eqref{eq:imitation-dynamics} redistributes the actions among the various communities without modifying the population state, namely $\y(t)=\x(t)\1$ remains constant. The second property is instead a preliminary convergence result that will play a crucial role in the next subsection. It says that, if Assumption \ref{assumption2} is satisfied, then  whenever the population state trajectory $\y(t)=\x(t)\1$ converges to some value $\bar \y$, the system state trajectory $\x(t)$ converges to the balanced equilibrium point $ \y^{\bullet}{\etav^\top}$. In the following, we prove these two facts.

First, we introduce some notation. For two actions $i,j $ in $\mc A$ and a state vector $\x$ in $\mc X$, let $$\Lambda_{ij}(\x):=\sum_{h,k\in\mc H}x_{ih}W_{hk}x_{jk}\,.$$ 
Also, we indicate with $\dot{\y}$ the vector field obtained by summing the right-hand side of (\ref{eq:imitation-dynamics}) with respect to $h$, namely}
\be\label{vector-field-y}
\dot{y}_i=\ds\sum_{h\in\mc H}\dot x_{ih}
=\ds\sum_{j\in\mc A}\Lambda_{ji}(\x)f_{ji}(\x\1)-\Lambda_{ij}(\x)f_{ij}(\x\1)\,.\ee
{If we plug in the right hand side of \eqref{vector-field-y} the system state trajectory $\x(t)$, we obtain the time derivative of the corresponding population state trajectory $\y(t)=\x(t)\1$.}
If the community network $\mc G$ is undirected, then 
$$\Lambda_{ji}(\x)=\sum_{h,k\in\mc H}x_{jh}W_{hk}x_{ik}=\sum_{h,k\in\mc H}x_{ih}W_{hk}x_{jk}=\Lambda_{ij}(\x)\,.$$
Consequently, we can rewrite \eqref{vector-field-y} as
\be\label{dot y}
\dot{y}_i=\ds\sum_{j\in\mc A}\Lambda_{ij}(\x)\big(f_{ji}(\x\1)-f_{ij}(\x\1)\big)\,.
\ee

We have the following result.

\begin{proposition}\label{prop-Z} Consider a population game  $(\mc A,r)$, an undirected connected community network $\mc G=(\mc H, \etav, W)$, and an imitation mechanism $\f$ satisfying Assumption \ref{assumption}.  Let $\x(t)$ be the solution of the imitation dynamics~(\ref{eq:imitation-dynamics}) with initial condition $\x(0)$ in $\xrnash$, and put $\y(t)=\x(t)\1$. Then, 
\begin{enumerate}
\item[(i)] ${\y}(t)=\y(0)$, for all $t\geq 0$;
\item[(ii)] $\x(t)\in\xrnash$, for all $t\geq 0$.
\end{enumerate}

\end{proposition}
\begin{IEEEproof}
{\bf (i) }{  Consider $\x$ in $\xrnash$ and put $\y=\x\1$. If $y_i, y_j>0$ we have that $r_i(\y)=r_j(\y)$. This fact implies that for such $\x$ the right hand side of~\eqref{dot y} is $0$ for every $i$, namely $\dot\y=0$. 
This says that given an initial condition $\x(0)$ in $\xrnash$, if we consider the affine subset $\mc X (\x(0)):=\{\x\in\mc X: \x\1=\x(0)\1\}\subseteq \xrnash$, we have that the vector field in ~\eqref{eq:imitation-dynamics} is along $\mc X (\x(0)$ and thus, by standard results on ODE's, the solution $\x(t)$ in $ \mc X (\x(0))$ for every $t$. This proves (i).}

{\bf(ii)} It follows from item (i) and the definition of $\xrnash$.
\end{IEEEproof}

{Next example shows that, if the community network is not undirected, the set $\xrnash$ is not in general invariant for the imitation dynamics.}
\begin{example}\label{ex:znotinv} 
Consider a population game $(\mc A, r)$ with binary action set $\mc A=\{0,1\}$ and reward functions $r_0(\y)=y_1$ and $r_1(\y)=1-y_1$. 
Let the community network $\mc G=(\mc H, \etav, W)$ consist of two communities $\mc H=\{a,b\}$, with $\etav =(1/2, 1/2)$, $W_{aa}=W_{bb}=W_{ba}=1$ and $W_{ab}=2$ (see Figure \ref{fig:topC}). Let the imitation mechanism be such that $f_{01}(\y)=1-y_1$ and $f_{10}(\y)=y_1$, thus satisfying Assumption~\ref{assumption}. 

The set $\xrnash$ consists of two points corresponding to the two restricted Nash equilibria when there is only one action played in the population and a segment corresponding to the Nash equilibrium $\y=(1/2, 1/2)$:
$$\begin{array}{lll}
\xrnash&=&\left\{\ds \left(\begin{matrix}0&0\\1/2&1/2\end{matrix}\right),\, \ds\left(\begin{matrix}1/2&1/2\\0&0\end{matrix}\right) \right \}\\
&&\bigcup\,\left\{\left(\begin{matrix}q&1/2-q\\1/2-q&q\end{matrix}\right)\right\}_{0\leq q\leq\frac12}.\end{array}
$$
For the initial state $$\x(0)=\left(\begin{matrix}0&1/2\\1/2&0\end{matrix}\right),$$
an explicit computation shows that $$\dot{y}_1(0)=\dot{x}_{1a}(0)+\dot{x}_{1b}(0)=-1/8\,.$$ 
Thus, for sufficiently small $t>0$, we have that $0<y_1(t)<1/2$ so that $\x(t)$ does not belong $\xrnash$.
\end{example}

{Our next result shows the preliminary convergence result presented above.

\begin{proposition}\label{prop:convbal} Consider a population game  $(\mc A,r)$, an undirected connected community network $\mc G=(\mc H, \etav, W)$, and an imitation mechanism $\f$ satisfying Assumptions~\ref{assumption} and~\ref{assumption2}. Let $\x(t)$ be the solution of the imitation dynamics~(\ref{eq:imitation-dynamics}). If  $$\lim_{t\to+\infty}\x\1(t)= \y^{\bullet}\,,$$ then $\y^{\bullet}$ is a restricted Nash equilibrium and
\be\label{convbal}\lim_{t\to+\infty}\x(t)=\y^{\bullet}{\etav^\top}.\ee
\end{proposition}}

\begin{IEEEproof}
{The fact that $\y^{\bullet}$ is a restricted Nash equilibrium follows from \eqref{dot y} and connectivity of the community network with analogous arguments as in the proof of Theorem \ref{teo:restpoints}. 

We now prove \eqref{convbal}.} For action $i$ in $\mc A$ is such that $y^{\bullet}_i=0$, we necessarily have that $x_{ih}(t)\to 0$ as $t\to +\infty$ and the limit relation is verified. Assume now that $y^{\bullet}_i>0$ and define, for every community $h$ in $\mc H$,
$$\ba{lll}u_{ih}\!\!&\!\!=\!\!&\ds\sum\limits_{j\in\mc A}\sum\limits_{k\in\mc H}\Big(x_{jh}W_{hk}x_{ik}f_{ji}(\y)-x_{ih}W_{hk}x_{jk}f_{ij}(\y)\Big)\\ &&
\!\!\!\!\ds-
\sum\limits_{j\in\mc A}\sum\limits_{k\in\mc H}\Big(x_{jh}W_{hk}x_{ik}f_{ji}(\bar \y)-x_{ih}W_{hk}x_{jk}f_{ij}(\bar \y)\Big).\ea$$
Let $\mu=f_{ij}(\bar \y)$ such that $j$ for which $y^{\bullet}_j>0$ (they are all equal and strictly positive because of Assumptions~\ref{assumption} and ~\ref{assumption2}). We rewrite~\eqref{eq:imitation-dynamics} as
\begin{equation}\label{eq:ode-mod}
\dot{x}_{ih}=\mu\left(\eta_h\sum\limits_{k\in\mc H}W_{hk}x_{ik}-x_{ih}\sum\limits_{k\in\mc H}W_{hk}\eta_k\right) +u_{ih}(t).
\end{equation}
Defining 
$$\zeta_{ih}=\frac{x_{ih}}{\eta_h},\quad \rho_{ih}(t)=\frac{u_{ih}(t)}{\eta_h},\quad Q_{hk}=\frac{W_{hk}\eta_k}{\sum\limits_{k'\in\mc H}W_{hk'}\eta_{k'}},$$
we rewrite \eqref{eq:ode-mod} as
$$
\dot{\zeta}_{i}= \mu(Q-I)\zeta_i+\rho_{i}(t)\,,
$$
where ${\zeta}_{i}$ and $\rho_i(t)$ are the vectors whose components are, respectively, $\zeta_{ih}$ and $\rho_{ih}(t)$. Define now ${\zeta}^{av}=(I-\1{\etav^\top})\zeta$ and notice that, 
since ${\etav^\top}Q={\etav^\top}$, the following relations hold
$$
\dot{\zeta}^{av}_{i}= \mu(Q-I)(I-\1{\etav^\top})\zeta^{av}_i+(I-\1{\etav^\top})\rho_{i}(t)\,.
$$
Since the matrix $(Q-I)(I-\1{\etav^\top})$ is asymptotically stable and $(I-\1{\etav^\top})\rho_{i}(t)$ is infinitesimal for $t\to +\infty$, it follows that also ${\zeta}^{av}(t)$ is infinitesimal. This implies that $x_{ih}(t)\to y^{\bullet}_i\eta_h$, for $t\to +\infty$, yielding the claim.
 \end{IEEEproof}

A simple consequence of this result concerns the dynamics inside the invariant set $\xrnash$.
\begin{corollary} Consider a population game  $(\mc A,r)$, an undirected connected community network $\mc G=(\mc H, \etav, W)$, and an imitation mechanism $\f$ satisfying Assumptions~\ref{assumption} and~\ref{assumption2}. Let $\x(t)$ be the solution of the imitation dynamics~\eqref{eq:imitation-dynamics} with initial condition $\x(0)$ in $\xrnash$. Then, $\lim_{t\to+\infty}\x(t)=\x(0)\1{\etav^\top}$.
\end{corollary}
\begin{IEEEproof}
It follows from item (i) of Proposition \ref{prop-Z} and from Proposition \ref{prop:convbal}.
\end{IEEEproof}
We conclude this subsection with the following technical result, whose utility will become apparent in the following subsection and whose proof is reported in Appendix \ref{sec:proof-lemma-boundary}. 
\begin{lemma}\label{lemma:boundary}
Consider a population game  $(\mc A,r)$, an undirected connected community network $\mc G=(\mc H, \etav, W)$, and an imitation mechanism $\f$ satisfying Assumption~\ref{assumption} . 
Let $\x^\bullet$ in $\xrnash\setminus\xnash$ be a system state associated to a restricted Nash equilibrium $\y^\bullet=\x^\bullet\1$ that is not a Nash equilibrium. Then, there exists $\eps>0$ such that 
for the imitation dynamics \eqref{eq:imitation-dynamics} it holds true that  (indicated as usual $\y=\x\1$) $\dot y_i>0$,
for every action $i$ in $\mc A$ such that $r_i(y^\bullet)>r_j(y^\bullet)$ for $j$ in $\mc S_{\y^\bullet}$ 
 and 
every system state $\x$ in $\mc X$ such that $||\x-\x^\bullet||<\eps$ and  $y_i>0$.
\end{lemma}

\subsection{Network imitation dynamics for potential population games}\label{sec:potential}

We now further specialize our analysis to the important  special case where the population game is potential. Under this assumption, and for community networks that are undirected and connected, we will prove global asymptotic convergence of network imitation dynamics to the set of Nash equilibria. 

We first recall the notion of potential~\cite{Monderer1996} in the context of continuous population games.  
\begin{definition}[Potential games]
A population game $(\mc A,r)$ is a potential population game if there exists a differentiable {potential function} $\Phi:\mc Y\to\R$ such that  
\be\label{eq:potential}r_j(\y)-r_i(\y)=\frac{\partial}{\partial y_j}\Phi(\y)-\frac{\partial}{\partial y_i}\Phi(\y)\,,\ee
for every actions $i,j$ in $\mc A$ and population state $\y$ in $\mc Y$.
\end{definition}

Potential games are a class of population games that include coordination and congestion games~\cite{Rosenthal1973, Monderer1996}, which have been extensively studied and used to model real-world phenomena such as the emergence of collective behaviors in social groups and traffic problems in infrastructure systems.

{It is known that for a potential population game all local maximum points of the potential function  $\Phi(\y)$ on $\mc Y$ are Nash equilibria and so are all internal stationary points of $\Phi(\y)$.

We now present a technical results that is key to our global convergence analysis. It shows that the potential function is a Lyapunov function for the imitation dynamics~\eqref{eq:imitation-dynamics}, as} it is always nondecreasing along the trajectories and stationary only on the set $\xrnash$. Its proof is reported in Appendix \ref{sec:proof-lemma-Lyapunov}.

\begin{lemma}\label{lemma:Lyapunov} Consider a potential population game  $(\mc A,r)$, an undirected connected community network $\mc G=(\mc H, \etav, W)$, and an imitation mechanism $\f$ satisfying Assumption \ref{assumption}. The derivative of the potential function $\Phi(\y)$ along the vector field of the imitation dynamics in \eqref{eq:imitation-dynamics} 
satisfies, for every $\x$ in $\mc X$ (indicated as usual $\y=\x\1$), 
\be\label{eq:Lyapunov}
\dot\Phi(\y)
=\ds\sum_{i\in\mc A}\frac{\partial \Phi(\y)}{\partial y_{i}}\dot y_{i}\geq 0.
\ee
Moreover, we have equality in \eqref{eq:Lyapunov} if and only if $\x$ belongs to $\xrnash$.  
\end{lemma}

A consequence of Lemma~\ref{lemma:Lyapunov}  is that every imitation dynamics in a potential population game with undirected community network has $\omega$-limit set\footnote{The $\omega$-limit set of a solution $x(t)$ of  \eqref{eq:imitation-dynamics} is the union of all the points $\bar\x$ such that there exists an increasing sequence of time instants $(t_k)_{k=1,\ldots}$ such that $\lim_{k\to+\infty}t_k=+\infty$ and $\lim_{k\to\infty}x(t_k)=\bar \x$.} contained in the set $\xrnash$.  In fact, combining this result with Lemma \ref{lemma:boundary} allows one to refine the characterization of such  $\omega$-limit set. 
Let 
$$\rnash_i=\{\y\in\mc \rnash:\,y_i=0\}\,,\qquad i\in\mc A\,,$$
and let 
$$\mc Y^{\circ}=\nash\cup\bigcup_{i\in\mc A\,:\rnash_i\cap\nash\ne\emptyset}\rnash_i$$
be the set containing all Nash equilibria and those restricted Nash equilibria lying on a face of the boundary of the simplex $\mc Y$ that contains a Nash equilibrium. We can now prove the following result.
\begin{theorem}\label{teo:potential}
Consider a potential population game  $(\mc A,r)$, an undirected connected community network $\mc G=(\mc H, \etav, W)$, and an imitation mechanism $\f$ satisfying Assumption~\ref{assumption}.  Then, for every initial state $\x(0)$ in $\mc X$ such that $\mc S_{\x(0)}=\mc A$, the solution $\x(t)$ of the imitation dynamics in \eqref{eq:imitation-dynamics} satisfies \footnote{The point to set distance is equal to $\dist(\x,{\mc X}^*):=\min_{\x'\in\mc X}||\x-\x'||$.}
$$\lim_{t\to+\infty}\dist(\x(t)\1,\mc Y^{\circ})=0\,.$$
\end{theorem}
\begin{IEEEproof} 
Put $\y(t)=\x(t)\1$. Since $\Phi(\y)$ is continuous on the compact set $\mc Y$, it is necessarily bounded. This and the fact that $\dot\Phi(\y(t))\geq 0$ by Lemma~\ref{lemma:Lyapunov} imply that  $\dot\Phi(\y(t))\to 0$ as $t\to +\infty$. Then, a continuity argument and the second part of Lemma~\ref{lemma:Lyapunov} imply that $\lim_{t\to+\infty}\dist(\y(t),\rnash)=0$. This implies that the $\omega$-limit set of $\y(t)$ is contained in a connected component $\mc C$ of the set of restricted Nash equilibria $\rnash$. For every action $i$ in $\mc A$, let $\mc C_i=
\mc C\cap\rnash_i$ be its intersection with the $i$-th face of the boundary of the simplex $\mc Y$. Now, recall that for every non-Nash point $\y^\bullet$ in $\rnash\setminus\nash$, there exists an action $i$ in $\mc A$ such that $y^\bullet_i=0$ and $r_i(\y^\bullet)>r_j(\y^\bullet)$, for $j$ in $\mc S_{\y^\bullet}$. It then follows from Lemma \ref{lemma:boundary} that, if $\mc C_i\cap\nash=\emptyset$, then $\dot y_i>0$ in an internal neighborhood of $\mc C_i$, so that no point in $\mc C_i$ is contained in the $\omega$-limit set of $\y(t)$. It then follows that the $\omega$-limit set of $\y(t)$ is contained in $\mc Y^{\circ}$. 
\end{IEEEproof}

Theorem~\ref{teo:potential} yields the following corollary guaranteeing global convergence of the imitation dynamics when the Nash equilibria of the population game are isolated { internal} points. 

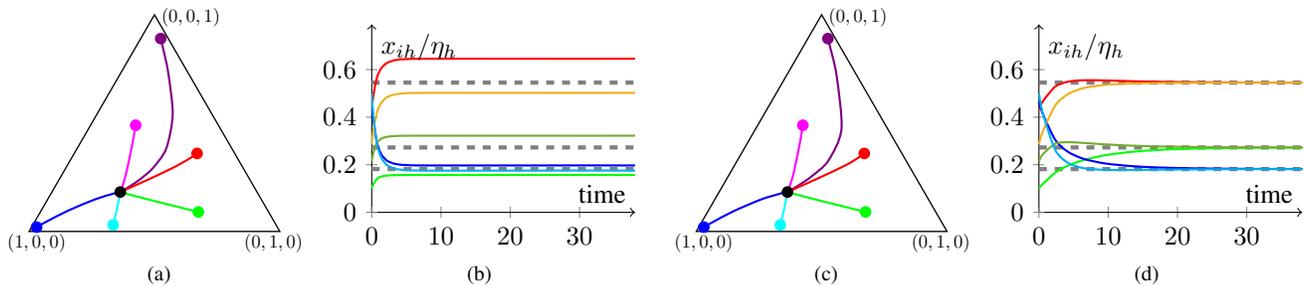
\begin{figure*}
\centering
\subfloat[]{\definecolor{mycolor2}{rgb}{0.00000,1.00000,1.00000}%
\definecolor{mycolor3}{rgb}{1.00000,0.00000,1.00000}%
\begin{tikzpicture}
\begin{axis}[
    width=4 cm,
scale only axis,
height=3.46 cm,
axis line style={draw=none},
ticks=none,
axis background/.style={fill=white},
]
\addplot[area legend, line width=0.5pt, draw=black, fill=white]
table[row sep=crcr] {%
x	y\\
0	0\\
1	0\\
0.5	0.866\\
0	0\\
}--cycle;

\node[draw=none] at (0.1cm,-.15cm) {\scalebox{0.7}{$(1,0,0)$}};
\node[draw=none] at (3.25cm,-.15cm) {\scalebox{0.7}{$(0,1,0)$}};
\node[draw=none] at (2.15cm,2.85cm) {\scalebox{0.7}{$(0,0,1)$}};

\addplot [color=violet, line width=0.8pt,smooth]
  table[row sep=crcr]{%
0.525	0.77076260936815\\
0.529353320917678	0.75294340628559\\
0.554114547830371	0.638641775814903\\
0.572674572030915	0.468924328695255\\
0.546872142675531	0.33105211025043\\
0.493047057063016	0.252629037345366\\
0.441358046687676	0.207652001943966\\
0.399699912772603	0.179232948713696\\
0.376128413277437	0.164861415405315\\
0.367325416921837	0.159655903847613\\
0.364471827013934	0.157965015761345\\
0.363785528751297	0.157550675112064\\
0.363664480130064	0.157476101584391\\
0.363641662775435	0.157462294754581\\
0.363637362362546	0.157459742787083\\
0.363636551868684	0.157459271217699\\
0.363636399113653	0.157459184078036\\
0.363636370323118	0.157459167975059\\
0.363636367283729	0.157459166331277\\
};

\addplot [color=violet, draw=none, mark=*, mark options={solid, fill=violet, violet}]
  table[row sep=crcr]{%
0.525	0.77076260936815\\
};

\addplot [color=blue, line width=0.8pt,smooth]
  table[row sep=crcr]{%
0.03	0.0173205080756887\\
0.0350927225449669	0.0202081041869684\\
0.0723814398447358	0.0408416079830512\\
0.154641200525795	0.0827259198033617\\
0.262918565812607	0.128166333259962\\
0.334538861010114	0.150565920082223\\
0.354816424959927	0.155488165830023\\
0.36113431012489	0.156907420434728\\
0.363088223441791	0.157337705808697\\
0.363540680260563	0.157437800746098\\
0.363618278717717	0.157455138547826\\
0.363632943925586	0.157458406553393\\
0.363635716941571	0.157459021720775\\
0.363636241339062	0.157459137489359\\
0.36363634050846	0.157459159274702\\
0.36363635917809	0.157459163347802\\
};

\addplot [color=blue, draw=none, mark=*, mark options={solid, fill=blue, blue}]
  table[row sep=crcr]{%
0.03	0.0173205080756887\\
};

\addplot [color=mycolor2, line width=0.8pt,smooth]
  table[row sep=crcr]{%
0.335	0.0259807621135331\\
0.348570884511006	0.0822296326930929\\
0.358882102638713	0.133355251587988\\
0.362453886135196	0.151457151271656\\
0.363360124349789	0.156058301675446\\
0.363589142781513	0.157219979767791\\
0.363627757537271	0.157415525998237\\
0.363634786789751	0.157451158053924\\
0.363636074442412	0.157457693877446\\
0.363636310591185	0.157458894206878\\
0.363636353906826	0.157459114702724\\
0.363636361851871	0.157459155208667\\
0.363636362919865	0.157459160667849\\
};

\addplot [color=mycolor2, draw=none, mark=*, mark options={solid, fill=mycolor2, mycolor2}]
  table[row sep=crcr]{%
0.335	0.0259807621135331\\
};

\addplot [color=red, line width=0.8pt,smooth]
  table[row sep=crcr]{%
0.67	0.311769145362398\\
0.657170098500997	0.302164299102617\\
0.582687181072481	0.258863841453469\\
0.481567640207949	0.210392086812152\\
0.411635942200499	0.178863087751561\\
0.380616109304989	0.165054864924198\\
0.369508145837741	0.160103155968337\\
0.365323622360299	0.158225808745258\\
0.364018314010501	0.157634462691521\\
0.363703885075554	0.157490280802461\\
0.363648945355018	0.157464885139017\\
0.36363870832647	0.157460215267306\\
0.363636800665015	0.157459357256145\\
0.363636445111026	0.157459199717794\\
0.363636378828832	0.157459170812459\\
0.363636366507315	0.157459165526596\\
};

\addplot [color=red, draw=none, mark=*, mark options={solid, fill=red, red}]
  table[row sep=crcr]{%
0.67	0.311769145362398\\
};

\addplot [color=green, line width=0.8pt,smooth]
  table[row sep=crcr]{%
0.675	0.0779422863405995\\
0.62765657016899	0.0895521373167568\\
0.475087225507023	0.127794827301997\\
0.408415272668413	0.145246069213155\\
0.382975627226325	0.152124483154403\\
0.370811516587931	0.155470789763416\\
0.365851432336566	0.156845427933855\\
0.364173171133741	0.157310854973325\\
0.363735824420772	0.157431770510449\\
0.36365549943512	0.157453873552935\\
0.363640040755152	0.157458143565856\\
0.363637070073544	0.157458967446026\\
0.363636499351353	0.157459126357234\\
0.363636389709083	0.157459157003476\\
0.363636368645374	0.157459162912941\\
0.363636364923747	0.157459163962638\\
};

\addplot [color=green, draw=none, mark=*, mark options={solid, fill=green, green}]
  table[row sep=crcr]{%
0.675	0.0779422863405995\\
};

\addplot [color=mycolor3, line width=0.8pt,smooth]
  table[row sep=crcr]{%
0.425	0.424352447854375\\
0.413243427391495	0.358105093740373\\
0.391303118274767	0.25082678924698\\
0.378892780531449	0.20278459475173\\
0.370675495824711	0.176513709829404\\
0.36612721661443	0.163837577483796\\
0.364262595503454	0.159028316996968\\
0.363753713492739	0.157753367041393\\
0.363657894978106	0.157514815094773\\
0.363640302264645	0.157469615231741\\
0.363637083026734	0.157461124496285\\
0.363636494863557	0.157459531912858\\
0.363636387542355	0.157459233260302\\
0.363636367985186	0.157459177253339\\
0.363636364426255	0.157459166749458\\
0.363636364118742	0.157459165790969\\
};

\addplot [color=mycolor3, draw=none, mark=*, mark options={solid, fill=mycolor3, mycolor3}]
  table[row sep=crcr]{%
0.425	0.424352447854375\\
};

\addplot [color=black, draw=none, mark=*, mark options={solid, fill=black, black}]
  table[row sep=crcr]{%
0.3636363636	0.1575\\
};

\end{axis}

\end{tikzpicture}\label{fig:type1}}\subfloat[]{\begin{tikzpicture}
\definecolor{mycolor1}{rgb}{0.00000,0.44700,0.74100}%
\definecolor{mycolor2}{rgb}{0.85000,0.32500,0.09800}%
\definecolor{mycolor3}{rgb}{1.00000,1.00000,0.00000}%
\definecolor{mycolor4}{rgb}{0.92900,0.69400,0.12500}%
\definecolor{mycolor5}{rgb}{0.49400,0.18400,0.55600}%
\definecolor{mycolor6}{rgb}{0.46600,0.67400,0.18800}%
\definecolor{mycolor7}{rgb}{0.30100,0.74500,0.93300}%
\definecolor{mycolor8}{rgb}{0.63500,0.07800,0.18400}%
\begin{axis}[
axis lines=middle,
 x   axis line style={->},
y   axis line style={->},
at={(0,0)},
    width=3.5 cm,
height=\h cm,
scale only axis,
xmin=0,
xmax=38,
xlabel={time},
 extra y ticks ={0},
    extra y tick labels={$0$},
     extra x ticks ={0},
    extra x tick labels={$0$},
ymin=0,
ymax=.79,
ylabel={$x_{ih}/\eta_h$},
axis background/.style={fill=white},
]

\addplot [ultra thick,dashed,color=gray]
  table[row sep=crcr]{%
0	0.181818181818\\
40	0.181818181818\\
};

\addplot [ultra thick,dashed,color=gray]
  table[row sep=crcr]{%
0	0.54545454545\\
40	0.54545454545\\
};

\addplot [ultra thick,dashed,color=gray]
  table[row sep=crcr]{%
0	0.2727272727\\
40	0.2727272727\\
};

\addplot [thick,smooth,color=red]
  table[row sep=crcr]{%
0	0.433333333333333\\
0.189457476696946	0.483504000571711\\
0.753403188927036	0.566378682977252\\
1.37666288752263	0.605688564307301\\
2.22177567838466	0.628466540725531\\
3.38087896443956	0.639972334344341\\
4.94827639408309	0.644458220914448\\
6.94335300914716	0.645659858713181\\
9.54889882125887	0.645881072990231\\
13.0770223713955	0.64589391648816\\
17.0770223713955	0.645865966014862\\
21.0770223713955	0.645821786937267\\
25.0770223713955	0.645755021341785\\
29.0770223713955	0.645661755274025\\
32.7240561882763	0.645702606403673\\
36.0129236833814	0.645823711594023\\
39.3052154551474	0.645882516353761\\
40	0.645909733513531\\
};

\addplot [thick,smooth,color=green]
  table[row sep=crcr]{%
0	0.1\\
0.189457476696946	0.115969360559676\\
0.753403188927036	0.140601000306159\\
1.37666288752263	0.150227919227549\\
2.22177567838466	0.154625734043733\\
3.38087896443956	0.156377470997867\\
4.94827639408309	0.156951700209164\\
6.94335300914716	0.157094325509032\\
9.54889882125887	0.15711989545083\\
13.0770223713955	0.157120365593317\\
17.0770223713955	0.157113455855282\\
21.0770223713955	0.157100506009284\\
25.0770223713955	0.157077141726299\\
29.0770223713955	0.157036803700887\\
32.7240561882763	0.157034178854506\\
36.0129236833814	0.157073783298575\\
39.3052154551474	0.157096508124479\\
40	0.157112314733282\\
};

\addplot [thick,smooth,color=blue]
  table[row sep=crcr]{%
0	0.466666666666667\\
0.189457476696946	0.400526638868613\\
0.753403188927036	0.293020316716589\\
1.37666288752263	0.24408351646515\\
2.22177567838466	0.216907725230736\\
3.38087896443956	0.203650194657792\\
4.94827639408309	0.198590078876388\\
6.94335300914716	0.197245815777787\\
9.54889882125887	0.196999031558938\\
13.0770223713955	0.196985717918523\\
17.0770223713955	0.197020578129856\\
21.0770223713955	0.197077707053449\\
25.0770223713955	0.197167836931916\\
29.0770223713955	0.197301441025088\\
32.7240561882763	0.197263214741821\\
36.0129236833814	0.197102505107401\\
39.3052154551474	0.19702097552176\\
40	0.196977951753188\\
};

\addplot [thick,smooth,color=mycolor4]
  table[row sep=crcr]{%
0	0.285714285714286\\
0.189457476696946	0.335647571668302\\
0.753403188927036	0.419954803629063\\
1.37666288752263	0.460461299727058\\
2.22177567838466	0.484106148674234\\
3.38087896443956	0.496127733740889\\
4.94827639408309	0.500835566564627\\
6.94335300914716	0.502099114237697\\
9.54889882125887	0.502331883806882\\
13.0770223713955	0.502345711707722\\
17.0770223713955	0.502317492793729\\
21.0770223713955	0.502273521516141\\
25.0770223713955	0.502208168685164\\
29.0770223713955	0.502118833389134\\
32.7240561882763	0.502161904967492\\
36.0129236833814	0.502279625279839\\
39.3052154551474	0.502335867152317\\
40	0.502360264989617\\
};

\addplot [thick,smooth,color=mycolor6]
  table[row sep=crcr]{%
0	0.214285714285714\\
0.189457476696946	0.24528649564916\\
0.753403188927036	0.291868434130295\\
1.37666288752263	0.309638091993948\\
2.22177567838466	0.317682196751321\\
3.38087896443956	0.320889008359177\\
4.94827639408309	0.32194577687676\\
6.94335300914716	0.322209238447302\\
9.54889882125887	0.322256546930323\\
13.0770223713955	0.322257645432543\\
17.0770223713955	0.322245755632649\\
21.0770223713955	0.32222368175524\\
25.0770223713955	0.322184106167781\\
29.0770223713955	0.322115949960076\\
32.7240561882763	0.32211068889539\\
36.0129236833814	0.3221762951294\\
39.3052154551474	0.322213954415399\\
40	0.322239957600703\\
};

\addplot [thick,smooth,color=cyan]
  table[row sep=crcr]{%
0	0.5\\
0.189457476696946	0.419065932682539\\
0.753403188927036	0.288176762240643\\
1.37666288752263	0.229900608278995\\
2.22177567838466	0.198211654574445\\
3.38087896443956	0.182983257899934\\
4.94827639408309	0.177218656558613\\
6.94335300914716	0.175691647315001\\
9.54889882125887	0.175411569262795\\
13.0770223713955	0.175396642859735\\
17.0770223713955	0.175436751573622\\
21.0770223713955	0.17550279672862\\
25.0770223713955	0.175607725147054\\
29.0770223713955	0.17576521665079\\
32.7240561882763	0.175727406137118\\
36.0129236833814	0.175544079590761\\
39.3052154551474	0.175450178432284\\
40	0.17539977740968\\
};

\end{axis}

\end{tikzpicture}\label{fig:configuration1}}\quad\subfloat[]{\definecolor{mycolor2}{rgb}{0.00000,1.00000,1.00000}%
\definecolor{mycolor3}{rgb}{1.00000,0.00000,1.00000}%
\begin{tikzpicture}
\begin{axis}[
    width=4 cm,
scale only axis,
height=3.46 cm,
axis line style={draw=none},
ticks=none,
axis background/.style={fill=white},
]
\addplot[area legend, line width=0.5pt, draw=black, fill=white]
table[row sep=crcr] {%
x	y\\
0	0\\
1	0\\
0.5	0.866\\
0	0\\
}--cycle;

\node[draw=none] at (0.1cm,-.15cm) {\scalebox{0.7}{$(1,0,0)$}};
\node[draw=none] at (3.25cm,-.15cm) {\scalebox{0.7}{$(0,1,0)$}};
\node[draw=none] at (2.15cm,2.85cm) {\scalebox{0.7}{$(0,0,1)$}};

\addplot [color=violet, line width=0.8pt,smooth]
  table[row sep=crcr]{%
0.525	0.77076260936815\\
0.529326733459215	0.753636277863564\\
0.559018720781618	0.626384013667154\\
0.579031789825307	0.408540428048419\\
0.52614670600482	0.282086852782739\\
0.460271315484123	0.220409890050623\\
0.412026527661054	0.187022924916426\\
0.379858581711105	0.167260210239377\\
0.368777977620883	0.160583592950918\\
0.364904160689369	0.158238888379232\\
0.363917085958551	0.157633231552582\\
0.363698516777534	0.157497873943636\\
0.363650124871659	0.157467755517961\\
0.363639410580166	0.157461069030154\\
0.363637038289812	0.157459586365326\\
0.363636690293545	0.157459368725213\\
};

\addplot [color=violet, draw=none, mark=*, mark options={solid, fill=violet, violet}]
  table[row sep=crcr]{%
0.525	0.77076260936815\\
};

\addplot [color=blue, line width=0.8pt,smooth]
  table[row sep=crcr]{%
0.03	0.0173205080756887\\
0.0344260103226957	0.0198470967829303\\
0.0673076736362377	0.0383239603036331\\
0.148796638734354	0.0810421208164906\\
0.260267960439961	0.128576629385704\\
0.336668071148017	0.151584627130011\\
0.3553118713074	0.155734756252918\\
0.361256086926586	0.156969523721606\\
0.363106676603657	0.157349379774671\\
0.363519196783622	0.157434782648176\\
0.363610434267957	0.157453757261903\\
0.363630623948563	0.157457966074111\\
0.363635092932004	0.157458898882003\\
0.363635985786954	0.157459085367405\\
};

\addplot [color=blue, draw=none, mark=*, mark options={solid, fill=blue, blue}]
  table[row sep=crcr]{%
0.03	0.0173205080756887\\
};

\addplot [color=mycolor2, line width=0.8pt,smooth]
  table[row sep=crcr]{%
0.335	0.0259807621135331\\
0.347214184375426	0.0752755784837131\\
0.358071659702698	0.129203126186184\\
0.362298380949931	0.150662888641295\\
0.363339416875858	0.15595248476276\\
0.363570722382498	0.157126331961217\\
0.363621841351258	0.157385556308571\\
0.363633149356993	0.157442875565041\\
0.363635652063078	0.157455558726574\\
0.363636206093342	0.157458366088536\\
0.363636328754345	0.157458987590611\\
};

\addplot [color=mycolor2, draw=none, mark=*, mark options={solid, fill=mycolor2, mycolor2}]
  table[row sep=crcr]{%
0.335	0.0259807621135331\\
};

\addplot [color=red, line width=0.8pt,smooth]
  table[row sep=crcr]{%
0.67	0.311769145362398\\
0.660531543248015	0.303709494923965\\
0.59890816174231	0.266154507866146\\
0.48961998697481	0.214081220073886\\
0.411670621989622	0.179068930867153\\
0.378822962492314	0.16434791440836\\
0.368446121734181	0.159657335916738\\
0.364829817771573	0.158008073310234\\
0.363900405862926	0.157581076130112\\
0.363694800858518	0.157486203537645\\
0.363649299617771	0.157465156943065\\
0.363639227568528	0.157460491906914\\
0.363636997733638	0.157459458366321\\
0.363636498592757	0.157459226917776\\
};

\addplot [color=red, draw=none, mark=*, mark options={solid, fill=red, red}]
  table[row sep=crcr]{%
0.67	0.311769145362398\\
};

\addplot [color=green, line width=0.8pt,smooth]
  table[row sep=crcr]{%
0.675	0.0779422863405995\\
0.632865193955099	0.0882990512547184\\
0.474417331478453	0.128035576003195\\
0.400529925060451	0.147400496977001\\
0.378508989680801	0.15336359471254\\
0.368812256857504	0.156028138358172\\
0.365098577038084	0.15705468201192\\
0.363960026411534	0.157369706849446\\
0.363707999585159	0.15743937287522\\
0.363652221246682	0.157454784150108\\
0.36363987435742	0.157458194708117\\
0.363637140931963	0.1574589496588\\
0.363636535741522	0.157459116795812\\
0.363636422742265	0.15745914800202\\
};

\addplot [color=green, draw=none, mark=*, mark options={solid, fill=green, green}]
  table[row sep=crcr]{%
0.675	0.0779422863405995\\
};

\addplot [color=mycolor3, line width=0.8pt,smooth]
  table[row sep=crcr]{%
0.425	0.424352447854375\\
0.414409226533035	0.360307487613572\\
0.38941038683568	0.238515543243913\\
0.376511946142713	0.192974531654938\\
0.369021529701378	0.171277923183303\\
0.365289249846711	0.161590716563347\\
0.36401328330795	0.158403570004954\\
0.363720309834862	0.157669612039385\\
0.363655011199736	0.157505915511334\\
0.3636405000165	0.157469534418428\\
0.363637280434437	0.157461462722401\\
0.363636566748744	0.157459673515054\\
0.363636408624102	0.157459277104823\\
0.363636380585624	0.157459206814538\\
};

\addplot [color=mycolor3, draw=none, mark=*, mark options={solid, fill=mycolor3, mycolor3}]
  table[row sep=crcr]{%
0.425	0.424352447854375\\
};

\addplot [color=black, draw=none, mark=*, mark options={solid, fill=black, black}]
  table[row sep=crcr]{%
0.3636363636	0.1575\\
};

\end{axis}

\end{tikzpicture}\label{fig:type2}}\subfloat[]{\begin{tikzpicture}
\definecolor{mycolor1}{rgb}{0.00000,0.44700,0.74100}%
\definecolor{mycolor2}{rgb}{0.85000,0.32500,0.09800}%
\definecolor{mycolor3}{rgb}{1.00000,1.00000,0.00000}%
\definecolor{mycolor4}{rgb}{0.92900,0.69400,0.12500}%
\definecolor{mycolor5}{rgb}{0.49400,0.18400,0.55600}%
\definecolor{mycolor6}{rgb}{0.46600,0.67400,0.18800}%
\definecolor{mycolor7}{rgb}{0.30100,0.74500,0.93300}%
\definecolor{mycolor8}{rgb}{0.63500,0.07800,0.18400}%
\begin{axis}[
axis lines=middle,
 x   axis line style={->},
y   axis line style={->},
at={(0,0)},
    width=3.5 cm,
height=\h cm,
scale only axis,
xmin=0,
xmax=38,
xlabel={time},
 extra y ticks ={0},
    extra y tick labels={$0$},
     extra x ticks ={0},
    extra x tick labels={$0$},
ymin=0,
ymax=.79,
ylabel={$x_{ih}/\eta_h$},
axis background/.style={fill=white},
]

\addplot [ultra thick,dashed,color=gray]
  table[row sep=crcr]{%
0	0.181818181818\\
40	0.181818181818\\
};

\addplot [ultra thick,dashed,color=gray]
  table[row sep=crcr]{%
0	0.54545454545\\
40	0.54545454545\\
};

\addplot [ultra thick,dashed,color=gray]
  table[row sep=crcr]{%
0	0.2727272727\\
40	0.2727272727\\
};

\addplot [thick,smooth,color=red]
  table[row sep=crcr]{%
0	0.433333333333333\\
0.560905318980092	0.466511290312438\\
2.35245354317924	0.528406247809314\\
4.14400176737838	0.548977629605953\\
6.41388392180545	0.555320036912128\\
9.51861168011581	0.554520608256259\\
13.4620273343614	0.551280582847612\\
17.4620273343614	0.548827647140176\\
21.4620273343614	0.547343826542411\\
25.4620273343614	0.546499750689652\\
29.4620273343614	0.546030012385245\\
33.4620273343614	0.54577078181006\\
37.4620273343614	0.545628194411608\\
40	0.545573233296347\\
};

\addplot [thick,smooth,color=green]
  table[row sep=crcr]{%
0	0.1\\
0.560905318980092	0.119191716410655\\
2.35245354317924	0.166511193886392\\
4.14400176737838	0.195642373987636\\
6.41388392180545	0.219478334541998\\
9.51861168011581	0.239844851755232\\
13.4620273343614	0.25466930888572\\
17.4620273343614	0.262848217445379\\
21.4620273343614	0.267312268129743\\
25.4620273343614	0.269756872851051\\
29.4620273343614	0.271097362495563\\
33.4620273343614	0.27183280310388\\
37.4620273343614	0.272236377448041\\
40	0.272391804773656\\
};

\addplot [thick,smooth,color=blue]
  table[row sep=crcr]{%
0	0.466666666666667\\
0.560905318980092	0.414296993276907\\
2.35245354317924	0.305082558304295\\
4.14400176737838	0.255379996406411\\
6.41388392180545	0.225201628545873\\
9.51861168011581	0.205634539988509\\
13.4620273343614	0.194050108266667\\
17.4620273343614	0.188324135414445\\
21.4620273343614	0.185343905327846\\
25.4620273343614	0.183743376459297\\
29.4620273343614	0.182872625119192\\
33.4620273343614	0.18239641508606\\
37.4620273343614	0.182135428140351\\
40	0.182034961929998\\
};

\addplot [thick,smooth,color=mycolor4]
  table[row sep=crcr]{%
0	0.285714285714286\\
0.560905318980092	0.339448230583572\\
2.35245354317924	0.449086751033917\\
4.14400176737838	0.496258381682544\\
6.41388392180545	0.522135863098376\\
9.51861168011581	0.535800177100026\\
13.4620273343614	0.54164028293544\\
17.4620273343614	0.543715432955889\\
21.4620273343614	0.544579654454449\\
25.4620273343614	0.544992138129943\\
29.4620273343614	0.545204711371741\\
33.4620273343614	0.545318305454015\\
37.4620273343614	0.545379967185061\\
40	0.545403619978162\\
};

\addplot [thick,smooth,color=mycolor6]
  table[row sep=crcr]{%
0	0.214285714285714\\
0.560905318980092	0.243754409493006\\
2.35245354317924	0.288214307959971\\
4.14400176737838	0.294864765926716\\
6.41388392180545	0.291843820403798\\
9.51861168011581	0.28577335801964\\
13.4620273343614	0.28022592408468\\
17.4620273343614	0.276907502353265\\
21.4620273343614	0.275036068630551\\
25.4620273343614	0.273997657250772\\
29.4620273343614	0.273425219708569\\
33.4620273343614	0.27311048703879\\
37.4620273343614	0.27293762766688\\
40	0.272871033875384\\
};

\addplot [thick,smooth,color=cyan]
  table[row sep=crcr]{%
0	0.5\\
0.560905318980092	0.416797359923422\\
2.35245354317924	0.262698941006111\\
4.14400176737838	0.20887685239074\\
6.41388392180545	0.186020316497826\\
9.51861168011581	0.178426464880333\\
13.4620273343614	0.17813379297988\\
17.4620273343614	0.179377064690847\\
21.4620273343614	0.180384276915001\\
25.4620273343614	0.181010204619286\\
29.4620273343614	0.18137006891969\\
33.4620273343614	0.181571207507194\\
37.4620273343614	0.181682405148059\\
40	0.181725346146454\\
};

\end{axis}

\end{tikzpicture}\label{fig:configuration2}}
\caption{Behavior of the imitation dynamics of Example~\ref{ex:ok} for the {pairwise proportional imitation} ((a) and (b)) and the  {sigmoid} imitation ((c) and (d)). In (a) and (c), different orbits of the population state with initial conditions plotted as colored circles. The imitation dynamics converges to the unique Nash equilibrium of the game (black circle). In (b) and (d), the time-evolution of the system state for the magenta orbit from (a) and (c), respectively, is compared with the balanced system state corresponding to the Nash equilibrium (the three horizontal dashed lines). The differently shaded red (green and blue) curves are the proportion of $0$-($1$- and $2$-)players in the two communities.  Parameters: $W_{aa}=W_{bb}=1$, $W_{ab}=W_{ba}=0.2$, $\etav=(0.7,0.3)$, and $K_{ij}=1$ for $i,j\in\mc A$.}
\label{fig:traj}
\end{figure*}

{
\begin{corollary}\label{cor:nash} Consider a potential population game  $(\mc A,r)$ with  set of Nash equilibria $\nash$ such that $\mc S_{\y^*}=\mc A$ for every $\y^*$ in $\nash$. Let $\mc G=(\mc H, \etav, W)$ be an undirected connected community network, and  $\f$ be an imitation mechanism satisfying Assumption~\ref{assumption}.  Then,
for every initial system state $\x(0)$ in $\mc X$ such that $\mc S_{\x(0)}=\mc A$, 
\be\label{Nashconv1}\lim_{t\to+\infty}\dist(\x(t)\1,{\nash})= 0\,.\ee
Moreover, if $\nash$ is finite, then there exists a Nash equilibrium $\y^*$ in $\nash$ such that 
\be\label{Nashcon2}\lim_{t\to+\infty}\x(t)\1=\y^*\,,\ee
and  if Assumption~\ref{assumption2} is also satisfied, then 
\be\label{Nashcon3}\lim_{t\to+\infty}\x(t)=\y^{*}{\etav^\top}\,.\ee
\end{corollary}
\begin{IEEEproof}
Since all Nash equilibria are in the interior of $\mc Y$, we have $\nash=\mc Y^{\circ}$, so that \eqref{Nashconv1} follows from Theorem~\ref{teo:potential}. When $\nash$ is finite, the Nash equilibria are isolated and \eqref{Nashconv1} implies \eqref{Nashcon2}.  Finally,  Proposition \ref{prop:convbal} and \eqref{Nashcon2} imply \eqref{Nashcon3} when Assumption~\ref{assumption2} is satisfied. 
 \end{IEEEproof}
 }

\begin{remark}
The assumption on the initial condition that $\mc S_{\x(0)}=\mc A$, i.e., $y_i(0)>0$, for all $i$ in $\mc A$ is not restrictive. Indeed, in the general case when some actions are not played at the beginning, by virtue of item (ii) of Proposition~\ref{proposition:imitations-properties}, we can restrict the game and the dynamics to those actions that are in the support of the initial condition $S_{\x(0)}$.
\end{remark}

We conclude this section by presenting two numerical examples: the first one validates the analytical predictions of Corollary~\ref{cor:nash}, while the second one shows that the result does not hold true in general when the game is not potential.

\begin{example}\label{ex:ok} {Consider a population game $(\mc A, r)$ with three actions $\mc A=\{0,1,2\}$ and reward functions given by
$$\ba{l}
r_0({y})=-2y_0,\quad r_1({y})=-4y_1,\quad\text{and}\quad r_2({y})=-6y_2.
\ea$$
This is a simple archetype of a congestion game \cite{Rosenthal1973} and is a potential game with potential function 
$$
\Phi(\y)=-y_0^2-2y_1^2-3y_2^2\,.
$$
Its unique Nash equilibrium is given by $\bar \y=\left(\frac{6}{11},\frac{3}{11},\frac{2}{11}\right)$, and coincides with the unique global maximum of $\Phi$.

Consider an} undirected connected community network $\mc G=(\mc H, \etav, W)$ consisting of two communities $\mc H=\{a,b\}$, as in Figure \ref{fig:topC}, with $W_{ab}=W_{ba}$. We study the behavior of two different imitation mechanisms: the pairwise proportional imitation in Example~\ref{ex:pairwise} and the {sigmoid} imitation rates in Example~\ref{ex:sigmoid}. Only the latter satisfies Assumption~\ref{assumption2}. Numerical simulations reported in Figure \ref{fig:traj} validate the analytical predictions of Corollary~\ref{cor:nash}. For both the dynamics, trajectories in the space of population states $\mc Y$ converge to the Nash equilibrium,  as shown in Figures~\ref{fig:type1} and~\ref{fig:type2}. On the other hand,  the behavior of the trajectories in the space of system states $\mc X$ is quite different in the two cases: for the pairwise proportional imitation, the system state trajectory converges to a system state that is not balanced (see Figure \ref{fig:configuration1});  for the {sigmoid} imitation dynamics, instead, consistently with item (ii) of Corollary~\ref{cor:nash}, the system state trajectory converges to the balanced system state (see Figure \ref{fig:configuration2}).
\end{example}

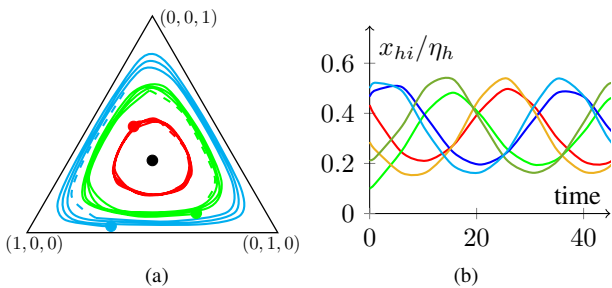
\begin{figure}
\centering
\subfloat[]{\begin{tikzpicture}
\begin{axis}[
    width=4 cm,
scale only axis,
height=3.46 cm,
axis line style={draw=none},
ticks=none,
axis background/.style={fill=white},
]
\addplot[area legend, line width=0.5pt, draw=black, fill=white]
table[row sep=crcr] {%
x	y\\
0	0\\
1	0\\
0.5	0.866\\
0	0\\
}--cycle;

\node[draw=none] at (0.1cm,-.15cm) {\scalebox{0.7}{$(1,0,0)$}};
\node[draw=none] at (3.25cm,-.15cm) {\scalebox{0.7}{$(0,1,0)$}};
\node[draw=none] at (2.15cm,2.85cm) {\scalebox{0.7}{$(0,0,1)$}};

\addplot [color=cyan, line width=0.8pt,smooth]
  table[row sep=crcr]{%
0.335	0.0259807621135331\\
0.28196904564268	0.0286050762489119\\
0.150738179549836	0.0517754951429989\\
0.121711867157312	0.114844575940001\\
0.169843680087855	0.238620981094432\\
0.306065379997874	0.484037448763552\\
0.440973637770194	0.681784519581083\\
0.507835560385198	0.71289382485236\\
0.581265424123774	0.650261736790647\\
0.71790447729392	0.437785923748138\\
0.852032705858232	0.180160643011447\\
0.849676585684742	0.0685212627004574\\
0.680325278330944	0.0327088717018892\\
0.264747584186061	0.0375902170927366\\
0.159074287489853	0.0695475963929857\\
0.145966003360388	0.147545631690755\\
0.223692075076285	0.322421730863987\\
0.378069100613567	0.580878455503279\\
0.474789129355282	0.683103319949546\\
0.560982063427826	0.652512375734797\\
0.680254892735853	0.484877373200991\\
0.821710379382105	0.222767325145367\\
0.831160420479562	0.083081538611617\\
0.648212399954033	0.0401603723932951\\
0.256947677457569	0.050611291412736\\
0.164764303443836	0.102553562370145\\
0.187746116410681	0.227761578614672\\
0.32080895296657	0.476319024077974\\
0.460324107339667	0.653509954218157\\
0.579126739784994	0.613816167941126\\
0.712750508200776	0.41506981080118\\
0.820320701880355	0.194527047471685\\
0.797537712228481	0.0816674502190093\\
0.562051042953935	0.0446400095553079\\
0.27506145779891	0.0587992590493561\\
};

\addplot [color=cyan, dashed, line width=0.8pt,smooth]
  table[row sep=crcr]{%
0.27506145779891	0.0587992590493561\\
0.177637510206542	0.148078940272131\\
0.244260116209395	0.326353642191358\\
0.38725591641293	0.560800961029322\\
};

\addplot [color=cyan, draw=none, mark=*, mark options={solid, fill=cyan, cyan}]
  table[row sep=crcr]{%
0.335	0.0259807621135331\\
};

\addplot [color=green, line width=0.8pt,smooth]
  table[row sep=crcr]{%
0.675	0.0779422863405995\\
0.641536457700728	0.072483528064156\\
0.432761505085286	0.0663391764394244\\
0.225426717612845	0.133513657150723\\
0.278421939429072	0.348554873307179\\
0.418122603232842	0.557769780674365\\
0.548906355732258	0.582814852383829\\
0.690343568738818	0.400718672106193\\
0.776063378672486	0.200505893518387\\
0.715980256159062	0.0966080592748151\\
0.426581068593375	0.0717665044361257\\
0.235457387314065	0.138294574161599\\
0.283852603029592	0.345703130967575\\
0.443676783385633	0.568497146415165\\
0.573859667157457	0.550877391168454\\
0.715344961367045	0.343496057728284\\
0.765438227265074	0.160027375390421\\
0.635424401563895	0.084131384435073\\
0.309276357199158	0.0969167638823427\\
0.243958892935213	0.233015251838358\\
0.360362518438482	0.464382656169644\\
0.497586886154534	0.578519324302776\\
0.635789657515578	0.46656732162894\\
0.757683358733027	0.211897554294758\\
0.672363832715914	0.0983744581688579\\
0.370974177261431	0.0890864480522507\\
0.246986642765272	0.213963661170896\\
0.35519824308458	0.447662412777639\\
0.488316166568235	0.568192164915826\\
};

\addplot [color=green, dashed, line width=0.8pt,smooth]
  table[row sep=crcr]{%
0.488316166568235	0.568192164915826\\
0.621009976032178	0.478072757012648\\
0.748488778351947	0.224335097208821\\
0.706503351725983	0.119414325074061\\
};

\addplot [color=green, draw=none, mark=*, mark options={solid, fill=green, green}]
  table[row sep=crcr]{%
0.675	0.0779422863405995\\
};

\addplot [color=red, line width=0.8pt,smooth]
  table[row sep=crcr]{%
0.425	0.424352447854375\\
0.450275229640249	0.444329849952901\\
0.566815912178567	0.427759965088504\\
0.648602210917994	0.297721890706239\\
0.6219089229498	0.178507476461194\\
0.462850114686276	0.150961207986794\\
0.343653580554481	0.239107075784615\\
0.402227095113498	0.393662216131088\\
0.507949639981347	0.456285109229395\\
0.620143665844083	0.359220984592499\\
0.653912388333624	0.227945198833806\\
0.530833956179924	0.151096046437715\\
0.361960729699016	0.195277238481624\\
0.358595104606816	0.313407534337567\\
0.443515545425404	0.434110139311033\\
0.551194317188095	0.437187598754504\\
0.639431878571083	0.317441271875547\\
0.635860188409199	0.194916001673287\\
0.492321984709922	0.150976648308794\\
0.350886516954892	0.218450199413123\\
0.375333273644178	0.347603154517968\\
0.47031611443888	0.447952286081789\\
0.574866114715171	0.415718769957898\\
0.648226623397375	0.287485120154555\\
0.603279059065406	0.171949955736895\\
0.438373543462561	0.157941324628543\\
0.347578617801063	0.261171609994237\\
0.413349417447503	0.40261668494184\\
0.522355755072246	0.449072370280768\\
};

\addplot [color=red, dashed, line width=0.8pt,smooth]
  table[row sep=crcr]{%
0.522355755072246	0.449072370280768\\
0.628153450781477	0.338340365894911\\
0.644887204029743	0.213601877839909\\
0.552863401391364	0.157204575750725\\
};

\addplot [color=red, draw=none, mark=*, mark options={solid, fill=red, red}]
  table[row sep=crcr]{%
0.425	0.424352447854375\\
};

\addplot [color=black, draw=none, mark=*, mark options={solid, fill=black, black}]
  table[row sep=crcr]{%
0.5	0.2887\\
};

\end{axis}

\end{tikzpicture}}\subfloat[]{\begin{tikzpicture}
\definecolor{mycolor1}{rgb}{0.00000,0.44700,0.74100}%
\definecolor{mycolor2}{rgb}{0.85000,0.32500,0.09800}%
\definecolor{mycolor3}{rgb}{1.00000,1.00000,0.00000}%
\definecolor{mycolor4}{rgb}{0.92900,0.69400,0.12500}%
\definecolor{mycolor5}{rgb}{0.49400,0.18400,0.55600}%
\definecolor{mycolor6}{rgb}{0.46600,0.67400,0.18800}%
\definecolor{mycolor7}{rgb}{0.30100,0.74500,0.93300}%
\definecolor{mycolor8}{rgb}{0.63500,0.07800,0.18400}%
\begin{axis}[
axis lines=middle,
 x   axis line style={->},
y   axis line style={->},
at={(0,0)},
    width=3.2 cm,
height=\h cm,
scale only axis,
xmin=0,
xmax=45,
xlabel={time},
 extra y ticks ={0},
    extra y tick labels={$0$},
     extra x ticks ={0},
    extra x tick labels={$0$},
ymin=0,
ymax=.75,
ylabel={$x_{hi}/\eta_h$},
axis background/.style={fill=white},
]

\addplot [thick,smooth,color=red]
  table[row sep=crcr]{%
0	0.433333333333333\\
1.11077761784314	0.386952442075683\\
6.05957350222101	0.243580381195165\\
10.6710971029629	0.211221754151873\\
15.6710971029629	0.273515514478072\\
20.6710971029629	0.409368250166791\\
25.6710971029629	0.497333015408727\\
30.6710971029629	0.429769253904771\\
35.1903733392637	0.29874720118674\\
40.1903733392637	0.206117662690082\\
44.7198058786985	0.203622434330993\\
50	0.297185607204074\\
};

\addplot [thick,smooth,color=green]
  table[row sep=crcr]{%
0	0.1\\
1.11077761784314	0.121096029239734\\
6.05957350222101	0.251730238311472\\
10.6710971029629	0.406429051328186\\
15.6710971029629	0.482324033160965\\
20.6710971029629	0.394985691767261\\
25.6710971029629	0.258789805463107\\
30.6710971029629	0.195178361002787\\
35.1903733392637	0.2185412172715\\
40.1903733392637	0.330080044195911\\
44.7198058786985	0.455738868305287\\
50	0.482836997093866\\
};

\addplot [thick,smooth,color=blue]
  table[row sep=crcr]{%
0	0.466666666666667\\
1.11077761784314	0.491951528684584\\
6.05957350222101	0.504689380493363\\
10.6710971029629	0.382349194519941\\
15.6710971029629	0.244160452360963\\
20.6710971029629	0.195646058065948\\
25.6710971029629	0.243877179128165\\
30.6710971029629	0.375052385092442\\
35.1903733392637	0.48271158154176\\
40.1903733392637	0.463802293114007\\
44.7198058786985	0.340638697363721\\
50	0.21997739570206\\
};

\addplot [thick,smooth,color=mycolor4]
  table[row sep=crcr]{%
0	0.285714285714286\\
1.11077761784314	0.253007256082504\\
6.05957350222101	0.161632346386415\\
10.6710971029629	0.165916355847932\\
15.6710971029629	0.27451922111913\\
20.6710971029629	0.460621169391783\\
25.6710971029629	0.539286597834391\\
30.6710971029629	0.411010131528555\\
35.1903733392637	0.249876340389238\\
40.1903733392637	0.165950676248617\\
44.7198058786985	0.185275451623065\\
50	0.323977375397217\\
};

\addplot [thick,smooth,color=mycolor6]
  table[row sep=crcr]{%
0	0.214285714285714\\
1.11077761784314	0.224874946270051\\
6.05957350222101	0.349042156703904\\
10.6710971029629	0.506833831109496\\
15.6710971029629	0.534979165494585\\
20.6710971029629	0.375079001255801\\
25.6710971029629	0.213191188838777\\
30.6710971029629	0.16258075192308\\
35.1903733392637	0.211273469894958\\
40.1903733392637	0.369349004332122\\
44.7198058786985	0.517633006259018\\
50	0.500049784662121\\
};

\addplot [thick,smooth,color=cyan]
  table[row sep=crcr]{%
0	0.5\\
1.11077761784314	0.522117797647445\\
6.05957350222101	0.489325496909681\\
10.6710971029629	0.327249813042572\\
15.6710971029629	0.190501613386284\\
20.6710971029629	0.164299829352416\\
25.6710971029629	0.247522213326832\\
30.6710971029629	0.426409116548365\\
35.1903733392637	0.538850189715804\\
40.1903733392637	0.464700319419261\\
44.7198058786985	0.297091542117917\\
50	0.175972839940662\\
};

\end{axis}

\end{tikzpicture}}
\caption{Behavior of the imitation dynamics of Example~\ref{ex:rps}. In (a), different orbits of the population state; in (b), time-evolution of the single entries of the system state corresponding to the red orbit in (a). Parameters: $W_{aa}=W_{bb}=1$, $W_{ab}=W_{ba}=0.2$, $\etav=(0.7,0.3)$, and $K_{ij}=1$, for $i,j\in\mc A$.}
\label{fig:rsp}
\end{figure}

\begin{example}\label{ex:rps} Consider the population game $(\mc A, r)$ with action set $\mc A=\{0,1,2\}$ and reward functions given by
$$\ba{l}
r_0({y})=y_1-y_2,\quad r_1({y})=y_2-y_0,\quad r_2({y})=y_0-y_1\,.
\ea$$
This game is the population version of the classical rock--paper--scissor game and is not potential (see details in~\cite[Chapter 3.3]{Sandholm2010}). The only Nash equilibrium is the population state $(1/3, 1/3, 1/3)$. We consider, as in previous example, an undirected connected community network $\mc G=(\mc H, \etav, W)$ consisting of two communities $\mc H=\{a,b\}$ with $W_{ab}=W_{ba}>0$. Figure \ref{fig:rsp} shows how, in this case, the trajectories of the {sigmoid} imitation dynamics (Example~\ref{ex:sigmoid}) exhibit an oscillating behavior and do not converge. Particularly, in (a) we plot different 
orbits of the population state, for diverse initial conditions, which show lack of convergence to the unique Nash equilibrium. The oscillatory behavior is even more visible in (b), when looking at the time-evolution of the single entries of the system state  corresponding.
\end{example}

\section{Conclusion}\label{sec:conclusion}

We have studied a novel deterministic model of imitation dynamics in population games over networks with community patterns{. The considered model allows one to account for, e.g., the presence of homophily in age, gender, and social groups}. Imitation dynamics are distributed learning mechanisms that rely on minimal information on the underlying game. Specifically, we have modeled the learning process through a system of nonlinear ordinary differential equations describing the evolution of the fraction of adopters of the different actions in each of the communities. 

Our main theoretical results are twofold. First, we have characterized the equilibrium points of the imitation dynamics, showing that the network plays a nontrivial role. Differently from the scenario without communities, where equilibria can be characterized in terms of Nash equilibria of the game, the network connectivity may determine the presence of other equilibrium points. Instead, for connected community networks, we demonstrate that the network structure may impose extra constraints on the feasible equilibria. Second, we have focused our analysis on potential games. For this class of games,  and when the community network is undirected and connected, we have proved global asymptotic convergence {and, under some further assumption on their structure, we have guaranteed that the  of the imitation dynamics converges to the set of Nash equilibria}. A number of examples have been discussed to validate our theoretical findings and show the role of the various assumptions in our statements. Our results contribute to expanding the state of the art in several directions: i) they provide a framework for studying imitation dynamics in networks with community patterns as opposed to fully mixed populations which are the case studied in much of the literature; ii) they ensure global stability of the Nash equilibria, whereas most of the literature is concerned with local stability; iii) they generalize the analysis to a broad class of learning dynamics that encompasses the replicator equation and other particular imitation dynamics considered in previous works.

Our results suggest various directions for future research. First, our analysis leaves some open problems concerning the characterization of the equilibrium points of the imitation dynamics and their stability when the network is directed and not connected, which should be addressed by future theoretical research. {Second, the extension of our theoretical results to the multi-population setting described in \cite{Sandholm2010} is an important avenue of future research. Third, the case of dynamic and adaptive topologies and communities should be explored. Fourth, toward a practical implementation of the proposed learning protocol in real-world scenarios, asynchronous communication protocols for imitation dynamics should be analyzed. Some preliminary results in this direction have been presented in~\cite{ecc2018} in a stochastic framework, where communication between players is temporized by random Poisson clock. However, a general theory for asynchronous imitation dynamics is still missing.} Finally, a case study should be proposed and analyzed, toward the application of our theoretical findings in real-world applications, such as traffic control~\cite{Como2013,Jiang2014} or planning of vaccination campaigns~\cite{Bauch2005}. 

 \appendix
 \subsection{Proof of Proposition \ref{proposition:imitations-properties}}\label{sect:proof-imitation-properties}
{\bf(i)} Define the vector fields $$F,G,H:\R^{\mc A\times\mc H}\to\R^{\mc A\times\mc H}$$ 
by putting, for every action $i$ in $\mc A$ and community $h$ in $\mc H$, 
$$F_{ih}(\x)=\sum_{j\in\mc A}\sum_{k\in\mc H}x_{jh}W_{hk}x_{ik}f_{ji}(\x\1),$$ 
$$G_{ih}(\x)=x_{ih}\sum_{j\in\mc A}\sum_{k\in\mc H}W_{hk}x_{jk}f_{ij}(\x\1),$$
$$H_{ih}(\x)=
\left\{\ba{lcl}
[F_{ih}(\x)]_+-G_{ih}(\x)&\se&x_{ih}\ge0,\\[3pt]
[F_{ih}(\x)]_+&\se&x_{ih}<0\,,
\ea\right.$$ 
for every $\x$ in $\R^{\mc A\times\mc H}$, 
{where $[a]_+=\max\{a,0\}$ stands for the positive part of a scalar $a$.}
Observe that  
{\be\label{F=G}
\ba{lcl}\ds\sum_{i\in\mc A}F_{ih}(\x)
&=&\ds\sum_{i\in\mc A}\sum_{j\in\mc A}\sum_{k\in\mc H}x_{jh}W_{hk}x_{ik}f_{ji}(\x\1)\\[7pt]
&=&\ds\sum_{j\in\mc A}\sum_{i\in\mc A}\sum_{k\in\mc H}x_{jh}W_{hk}x_{ik}f_{ji}(\x\1)\\[7pt]
&=&\ds\sum_{j\in\mc A}G_{jh}(\x)\,.\ea\ee}
Since $F(\x)$ and $G(\x)$ are Lipschitz-continuous on $\R^{\mc A\times\mc H}$ and $G_{ih}(\x)=0$ whenever $x_{ih}=0$, 
we get that $H(\x)$ is Lipschitz-continuous on $\R^{\mc A\times\mc H}$. Hence, the dynamical system \be\label{H-system}\dot \x=H(\x)\ee admits a unique solution $(\x(t))_{t\ge0}$ for every  initial condition $\x(0)$ in $\R^{\mc A\times\mc H}$. 
Now, notice that $H_{ih}(\x)\ge0$ for $x_{ih}\le0$, so that whenever the initial condition has nonnegative entry $x_{ih}(0)\ge0$, the corresponding entry $x_{ih}(t)$ of the solution of \eqref{H-system} remains nonnegative for all $t\ge0$. I.e., the nonnegative orthant $\R_{+}^{\mc A\times\mc H}$ is invariant for \eqref{H-system}. On the other hand, for every $\x$ in $\R_{+}^{\mc A\times\mc H}$, we have that $F_{ih}(\x)\ge0$, so that 
\be\label{eq:H=F}H_{ih}(\x)=[F_{ih}(\x)]_+-G_{ih}(\x)=F_{ih}(\x)-G_{ih}(\x)\,,\ee
which, together with \eqref{F=G}, implies that 
$$\sum_{i\in\mc A}H_{ih}(\x)
=\sum_{i\in\mc A}\left(F_{ih}(\x)-G_{ih}(\x)\right)=0\,.$$ 
This proves that the set $\mc X$ is invariant for the system \eqref{H-system}. Finally, notice that it follows from \eqref{eq:H=F} that the system \eqref{H-system} coincides with \eqref{eq:imitation-dynamics} on $\mc X$ thus proving that the latter admits a unique solution for every initial condition $\x(0)$ in $\mc X$. 
 
{\bf(ii)} Observe from~\eqref{eq:imitation-dynamics} that, if $x_{ih}=0$ for all $h$ in $\mc H$, then $\dot x_{ih}=0$. Together with uniqueness of the solution, this implies that every solution of~\eqref{eq:imitation-dynamics} with $x_{ih}(0)=0$ for every $h$ in $\mc H$ is such that $x_{ih}(t)=0$ for every $h$ in $\mc H$ and $t\geq0$. On the other hand, for every initial condition $\x(0)$ in $\mc X$, the corresponding solution of~\eqref{eq:imitation-dynamics} satisfies the inequality $\dot x_{ih}(t)\ge-C_{i}x_{ih}(t)$, where $C_i=\max\{f_{ij}(\y):\,{\y\in\mc Y},j\in\mc A\}\,.$ Then, Gronwall's inequality implies that $$x_{ih}(t)\ge x_{ih}(0)e^{-C_it}>0\,,\qquad\forall\, t\geq 0\,,$$ 
which implies the claim.  \qed

{\subsection{Proof of Lemma \ref{lemma:boundary}}
\label{sec:proof-lemma-boundary}
For $\x^\bullet$ in $\xrnash\setminus\xnash$, and $\y^\bullet=\x^\bullet\1$, let $i$ in $\mc A$ be an action such that $y^\bullet_i=0$, and $r_i(\y^{\bullet})>r_j(\y^{\bullet})$ for every action $j$ such that $y^\bullet_j>0$. Then, on the one hand, by Assumption \ref{assumption}, 
$$f_{ji}(\y^{\bullet})-f_{ij}(\y^{\bullet})>0\,,\qquad\forall\,j\in\mc S_{\y^\bullet}
$$
and on the other hand, {there exist two positive constants $0<K_1<K_2<\infty$ such that
$$K_1y_i\leq\sum_{j\in\mc S_{\y^\bullet}}\Lambda_{ij}(\x)\leq K_2 y_i,\quad\text{ as }\quad \x\to\x^{\bullet}\,.$$
In contrast, for every $\ell$ that does not belong to $\mc S_{\y^\bullet}$, we have that there exists a positive constant $0<K_3<\infty$ such that
$$\Lambda_{i\ell}(\x)=\Lambda_{\ell i}(\x)\leq K_3 y_iy_\ell \quad\text{ as }\quad \x\to\x^{\bullet}\,.$$
Then, it follows from \eqref{dot y} that $$\ba{lll}
\dot{y}_i
&=&\ds\sum_{j\in\mc S_{\y^\bullet}}\Lambda_{ij}(\x)\big(f_{ji}(\x\1)-f_{ij}(\x\1)\big)\\
&&\ds+\sum_{k\notin\mc S_{\y^\bullet}}\Lambda_{ik}(\x)\big(f_{ki}(\x\1)-f_{ik}(\x\1)\big)\,\geq\, K_4 y_i\,,\ea$$
for some $K_4>0$,} as $\x\to\x^{\bullet}$. This implies that there exists $\eps>0$ such that $\dot y_i>0$ for every system state $\x$ in $\mc X$ such that $||\x-\x^\bullet||<\eps$ and  $y_i>0$, thus proving the claim.
\qed}

\subsection{Proof of Lemma \ref{lemma:Lyapunov}}
\label{sec:proof-lemma-Lyapunov}
Using \eqref{dot y} and \eqref{eq:potential}, we get 
\begin{equation}\label{eq:chain}
\begin{array}{l}
\dot\Phi(\y)
=\ds\sum_{i,j\in\mc A}\!\Lambda_{ij}(\x)\big(f_{ji}(\y)-f_{ij}(\y)\big)\frac{\partial \Phi(\y)}{\partial y_{i}}\\[7pt]
\,\,\,=\ds\frac12\sum_{i,j\in\mc A}\!\Lambda_{ij}(\x)\big(f_{ji}(\y)-f_{ij}(\y)\big)\ds\!\!\left(\frac{\partial \Phi(\y)}{\partial y_{i}}-\frac{\partial \Phi(\y)}{\partial y_{j}}\right)\\[7pt]
\,\,\,=\ds\frac12\sum_{i,j\in\mc A}\!\Lambda_{ij}(\x)\big(f_{ji}(\y)-f_{ij}(\y)\big)\big(r_i(\y)-r_j(\y)\big)\geq0,
\end{array}
\end{equation}
where the last inequality follows from Assumption~\ref{assumption} and the non-negativity of the term $\Lambda_{ij}(\x)$. This proves \eqref{eq:Lyapunov}.

On the other hand, \eqref{eq:chain} and Assumption~\ref{assumption} imply that we have equality in~\eqref{eq:Lyapunov} if and only if 
$$\Lambda_{ij}(\x)(r_i(\y)-r_j(\y))=0,\quad \forall\,i,j\in\mc A\,.$$
This implies that $r_i(\y)=r_j(\y)$ whenever $\Lambda_{ij}(\x)>0$, namely whenever the two actions $i,j$ in $\mc S_{\x}$ are played in the same community or in two communities connected by a link. 
Now, for every actions $i,j$ in $\mc S_{\x}$, there exist communities $h,k$ in $\mc H$ such that $x_{ih}>0$ and $x_{jk}>0$. Since the community network $\mc G$ is connected, it contains a path $h=h_0,h_1,\dots ,h_\ell=k$ from $h$ to $k$. Pick arbitrarily actions $j_1,\dots , j_{\ell-1}$ in $\mc A$ such that $x_{j_sh_s}>0$, for every $s=1,\dots , \ell-1$. Notice now that $$\Lambda_{ij_1}(\x)>0,\; \Lambda_{j_1j_2}(\x)>0,\dots , \Lambda_{j_{\ell-1}j}(\x)>0\,.$$
This implies that $r_i(\x\1)= r_{j_1}(\x\1)=\cdots = r_j(\x\1),$ which yields the claim. \qed

\begin{IEEEbiography}
[{\includegraphics[width=1in,height=1.25in,clip,keepaspectratio]{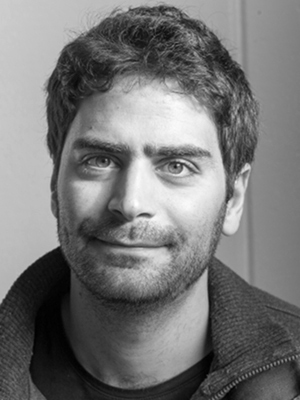}}]
{Giacomo Como}
is  an  Associate  Professor  at  the Department  of  Mathematical  Sciences,  Politecnico di  Torino,  Italy,  and  at  the  Automatic  Control  Department  of  Lund  University,  Sweden.  He  received the B.Sc., M.S., and Ph.D. degrees in Applied Mathematics  from  Politecnico  di  Torino,  in  2002,  2004, and 2008, respectively. He was a Visiting Assistant in  Research  at  Yale  University  in  2006--2007  and  a Postdoctoral  Associate  at  the  Massachusetts  Institute of Technology, from 2008 to 2011. He currently serves  as  Associate  Editor  of the  \textit{IEEE Transactions on Network Science and Engineering} and of the \textit{IEEE Transactions on Control of Network Systems} and  as  chair  of the  {IEEE-CSS  Technical  Committee  on  Networks  and  Communications}.  He was  the  IPC  chair  of  the  IFAC  Workshop  NecSys 2015  and  a  semiplenary speaker  at  the  International  Symposium  MTNS 2016 and the SICE ISCS 2017.  He  is  recipient  of  the 2015  George S.~Axelby  Outstanding  Paper award.  His  research interests  are in  dynamics,  information,  and  control  in  network  systems  with  applications to  cyber-physical  systems,  infrastructure  networks,  and  social  and  economic networks.
\end{IEEEbiography}

\begin{IEEEbiography}
[{\includegraphics[width=1in,height=1.25in,clip,keepaspectratio]{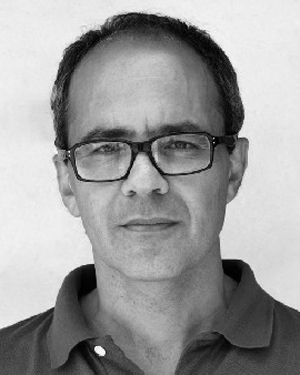}}]
{Fabio Fagnani}
received the Laurea degree in Mathematics from the University of Pisa and the Scuola Normale Superiore of Pisa,  Italy, in 1986. He received the PhD degree in Mathematics from the University of Groningen,  The  Netherlands,  in 1991. From 1991 to 1998, he was an Assistant Professor at the Scuola Normale Superiore. In 1997, he was a Visiting Professor at the Massachusetts Institute of Technology. Since 1998, he has been with the Politecnico of Torino, where  he has been a Full Professor of Mathematical Analysis since 2002. From 2006 to 2012, he has acted as a coordinator of the PhD program Mathematics for Engineering Sciences and  from 2012 to 2019 he was the head of the Department of Mathematical Sciences, Politecnico di Torino. He is an Associate Editor of the \textit{IEEE Transactions on Automatic Control} and served in the same role for the \textit{IEEE Transactions on Network Science and Engineering} and the \textit{IEEE Transactions on Control of Network Systems}. His current research topics are on cooperative algorithms and dynamical systems over graphs, inferential distributed algorithms, and opinion dynamics. 
\end{IEEEbiography}

\begin{IEEEbiography}
[{\includegraphics[width=1in,height=1.25in,clip,keepaspectratio]{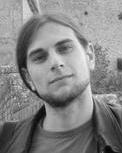}}]
{Lorenzo Zino}
has been a PostDoc Researcher at the University of Groningen, The Netherlands, since 2019. He received the B.Sc. in Applied Mathematics and the  M.S. in Mathematical Modeling from Politecnico di Torino, in 2012 and 2014, respectively, and the Ph.D. in Pure and Applied Mathematics jointly from Politecnico and Universit\`a di Torino, in 2018. He was a Research Fellow at Politecnico di Torino and a Visiting Research Assistant at New York University Tandon School of Engineering. His research interests include control of network systems, applied probability, network analysis, and game theory.
\end{IEEEbiography}

\end{document}